\def\compileforpublish{1}
\def\isaccepted{1}
\documentclass[journal,twoside]{IEEEtran}  %

\IEEEoverridecommandlockouts                              %

\usepackage{graphicx}
\graphicspath{ {./figures/} }
\usepackage[table]{xcolor}
\usepackage{calc}
\usepackage{tubscolors}
\usepackage{changes}
\usepackage{tikz}
\usetikzlibrary{shapes,
				automata,
				arrows,
				arrows.meta,
				matrix,
				backgrounds,
				fit,
				patterns,
				decorations.markings, 
				svg.path, 
				shapes.multipart, 
				shapes.geometric, 
				external, 
				shadows, 
				patterns,
				positioning, 
				calc, 
				decorations, 
				scopes,
				fadings,
				bending,
				scopes}
\usepackage{pgfplots}
\pgfplotsset{compat=1.17} 
\usepackage{pgfplotstable}
\usepackage{enumitem}
\newlist{RQ}{enumerate}{1}
\setlist[RQ]{	label=\emph{\textbf{RQ\,\arabic*:}}, %
				ref={RQ\,\arabic*}, %
				align=parleft, %
				left=\parindent, %
				leftmargin=*, %
				itemsep=3pt, %
				before={\itshape} %
			}
\usepackage{siunitx}
\usepackage[%
			bibstyle=ieee,
			citestyle=numeric-comp,
			backend=biber,
			minnames=1,
			maxcitenames=2,
			maxbibnames=6,
			isbn=false,
			url=false,
			natbib=true,
			clearlang=true,
			date=year
			]{biblatex} 
\usepackage{xpatch}
\usepackage{xstring}

\DeclareSourcemap{
	\maps[datatype=bibtex]{
		\map{
			\step[fieldsource=langid, match=\regexp{\A(n)?((swiss)?german|austrian)\Z}, final]
			\step[fieldset=language, fieldvalue={(in German)}]
		}
		\map{
			\step[fieldsource=langid, match=pinyin, final]
			\step[fieldset=language, fieldvalue={(in Chinese)}]
		}
		\map{
			\step[fieldsource=langid, match=japanese, final]
			\step[fieldset=language, fieldvalue={(in Japanese)}]
		}
		\map{
			\step[fieldsource=langid, match=ko, final]
			\step[fieldset=language, fieldvalue={(in Korean)}]
		}
	}
}

\xpatchbibdriver{inproceedings}					 	%
	{\usebibmacro{publisher+location+date}}			%
	{\IfSubStr{\strfield{booktitle}}{\strfield{year}}{\usebibmacro{publisher+location}}{\usebibmacro{publisher+location+date}}} %
	{} %
	{\typeout{There was an error patching biblatex-ieee (specifically, ieee.bbx's @inproceedings driver)}} %

\newbibmacro*{publisher+location}{%
	\printlist{location}%
	\iflistundef{publisher}
	{\setunit*{\addcomma\space}}
	{\setunit*{\addcolon\space}}%
	\printlist{publisher}%
	\newunit}

\xpatchbibdriver{online}
{\printtext[parens]{\usebibmacro{date}}}
	{\iffieldundef{year}
		{}
		{\printtext[parens]{\usebibmacro{date}}}}
	{}
	{\typeout{There was an error patching biblatex-ieee (specifically, ieee.bbx's @online driver)}}

\DeclareSourcemap{
	\maps[datatype=biber]{
		\map{
			\step[fieldsource=note, final]
			\step[fieldset=addendum, origfieldval, final]
			\step[fieldset=note, null]
		}
	}
}

\DeclareBibliographyDriver{standard}{%
	\usebibmacro{bibindex}%
	\usebibmacro{begentry}%
	\usebibmacro{maintitle+title}%
	\setunit{\addcomma\addspace}%
	\usebibmacro{publisher+type+number}%
	\setunit{\labelnamepunct}\newblock
	\IfSubStr{\strfield{number}}{\strfield{year}}{}{\usebibmacro{date}} %
	\newunit\newblock
	\usebibmacro{finentry}%
}

\newbibmacro*{publisher+type+number}{%
	\printtext{%
		\printlist{publisher}
		\iflistundef{publisher}
		{\space}{}%
		\iffieldundef{type}
		{Standard}
		{\printfield{type}}
		\printfield{number}
	}%
}

\xpatchbibdriver{report}					 	%
	{\usebibmacro{date}}			%
	{\IfSubStr{\strfield{number}}{\strfield{year}}{}{\usebibmacro{date}}} %
	{} %
	{\typeout{There was an error patching biblatex-ieee (specifically, ieee.bbx's @report driver)}} %

\DeclareBibliographyDriver{software}{%
	\usebibmacro{bibindex}%
	\usebibmacro{begentry}%
	\usebibmacro{maintitle+title}%
	\setunit{\addcomma\addspace}%
	\usebibmacro{version+year}%
	\setunit{\labelnamepunct}\newblock
	\usebibmacro{author}%
	\setunit{\addcomma\addspace}%
	\newunit\newblock
	\usebibmacro{url+urldate}%
	\newunit\newblock
	\usebibmacro{finentry}%
}

\newbibmacro*{version+year}{%
	\printtext{%
		\iffieldundef{version}{\iffieldundef{year}{}{\printfield{year}}}{\printfield{version}}
		\printfield{number}
		\printlist{publisher}
		\iflistundef{publisher}{\space}{}%
	}%
}

\DeclareSourcemap{
	\maps[datatype=bibtex]{
		\map[overwrite, foreach={booktitle,journaltitle,eventtitle,series,publisher,institution,type}]{
			\step[fieldsource=\regexp{$MAPLOOP}, match={XXX}, replace={Acad. Serbe Sci. Arts Glas Cl. Sci. Tech.}]
			\step[fieldsource=\regexp{$MAPLOOP}, match={XXX}, replace={Acta Acust.}]
			\step[fieldsource=\regexp{$MAPLOOP}, match={XXX}, replace={Acta Astron. (Poland)}]
			\step[fieldsource=\regexp{$MAPLOOP}, match={XXX}, replace={Acta Astron. Sin. (China)}]
			\step[fieldsource=\regexp{$MAPLOOP}, match={XXX}, replace={Acta Astronaut. (U.K.)}]
			\step[fieldsource=\regexp{$MAPLOOP}, match={XXX}, replace={Acta Astrophys. Sin. (China)}]
			\step[fieldsource=\regexp{$MAPLOOP}, match={XXX}, replace={Acta Autom. Sin.}]
			\step[fieldsource=\regexp{$MAPLOOP}, match={XXX}, replace={Acta Cienc. Indica Math.}]
			\step[fieldsource=\regexp{$MAPLOOP}, match={XXX}, replace={Acta Cienc. Indica Phys.}]
			\step[fieldsource=\regexp{$MAPLOOP}, match={XXX}, replace={Acta Crystallogr. A, Found. Crystallogr.}]
			\step[fieldsource=\regexp{$MAPLOOP}, match={XXX}, replace={Acta Crystallogr. B, Struct. Sci.}]
			\step[fieldsource=\regexp{$MAPLOOP}, match={XXX}, replace={Acta Crystallogr. C, Cryst. Struct. Commun.}]
			\step[fieldsource=\regexp{$MAPLOOP}, match={XXX}, replace={Acta Electron. (France)}]
			\step[fieldsource=\regexp{$MAPLOOP}, match={XXX}, replace={Acta Cyberno}]
			\step[fieldsource=\regexp{$MAPLOOP}, match={XXX}, replace={Acta Electron. Sin. (China)}]
			\step[fieldsource=\regexp{$MAPLOOP}, match={XXX}, replace={Acta Geod. Geophys. Montan. Hung.}]
			\step[fieldsource=\regexp{$MAPLOOP}, match={XXX}, replace={Acta Geophys. Pol.}]
			\step[fieldsource=\regexp{$MAPLOOP}, match={XXX}, replace={Acta Geophys. Sin. (China)}]
			\step[fieldsource=\regexp{$MAPLOOP}, match={XXX}, replace={Acta Geophys. Sin. (USA)}]
			\step[fieldsource=\regexp{$MAPLOOP}, match={XXX}, replace={Acta Metall.}]
			\step[fieldsource=\regexp{$MAPLOOP}, match={XXX}, replace={Acta Mex. Cienc. Tecnol.}]
			\step[fieldsource=\regexp{$MAPLOOP}, match={XXX}, replace={Acta Phys. Hung.}]
			\step[fieldsource=\regexp{$MAPLOOP}, match={XXX}, replace={Acta Phys. Pol. A}]
			\step[fieldsource=\regexp{$MAPLOOP}, match={XXX}, replace={Acta Phys. Pol. B}]
			\step[fieldsource=\regexp{$MAPLOOP}, match={XXX}, replace={Acta Phys. Sin.}]
			\step[fieldsource=\regexp{$MAPLOOP}, match={XXX}, replace={Acta Phys. Slovaca}]
			\step[fieldsource=\regexp{$MAPLOOP}, match={XXX}, replace={Acta Politec. Mex.}]
			\step[fieldsource=\regexp{$MAPLOOP}, match={XXX}, replace={Acta Polytech. Scand. Appl. Phys. Ser.}]
			\step[fieldsource=\regexp{$MAPLOOP}, match={XXX}, replace={Acta Polytech. Scand. Chem. Technol.}]
			\step[fieldsource=\regexp{$MAPLOOP}, match={XXX}, replace={Metall. Ser.}]
			\step[fieldsource=\regexp{$MAPLOOP}, match={XXX}, replace={Acta Polytech. Scand. Electr. Eng. Ser.}]
			\step[fieldsource=\regexp{$MAPLOOP}, match={XXX}, replace={Acta Polytech. Scand. Math. Comput. Sci. Ser.}]
			\step[fieldsource=\regexp{$MAPLOOP}, match={XXX}, replace={Acta Polytech. Scand. Mech. Eng. Ser.}]
			\step[fieldsource=\regexp{$MAPLOOP}, match={XXX}, replace={Acta Seismol. Sin.}]
			\step[fieldsource=\regexp{$MAPLOOP}, match={XXX}, replace={Acta Tech. Acad. Sci. Hung.}]
			\step[fieldsource=\regexp{$MAPLOOP}, match={XXX}, replace={Acta Tech. CSAV}]
			\step[fieldsource=\regexp{$MAPLOOP}, match={XXX}, replace={Acustica}]
			\step[fieldsource=\regexp{$MAPLOOP}, match={XXX}, replace={(AEU) Arch. Elektr. Ubertragung}]
			\step[fieldsource=\regexp{$MAPLOOP}, match={XXX}, replace={Akust. Zh.}]
			\step[fieldsource=\regexp{$MAPLOOP}, match={XXX}, replace={Algorithmica}]
			\step[fieldsource=\regexp{$MAPLOOP}, match={XXX}, replace={Alta Freq.}]
			\step[fieldsource=\regexp{$MAPLOOP}, match={XXX}, replace={An. Acad. Bras. Cienc.}]
			\step[fieldsource=\regexp{$MAPLOOP}, match={XXX}, replace={An. Fis.}]
			\step[fieldsource=\regexp{$MAPLOOP}, match={XXX}, replace={An. Mec. Electr.}]
			\step[fieldsource=\regexp{$MAPLOOP}, match={XXX}, replace={Angew. Inform.}]
			\step[fieldsource=\regexp{$MAPLOOP}, match={XXX}, replace={Ann. Inst. Henri Poincare Phys. Theor.}]
			\step[fieldsource=\regexp{$MAPLOOP}, match={XXX}, replace={Ann. Soc. Sci. Brux. I, Sci. Math. Astron. Phys.}]
			\step[fieldsource=\regexp{$MAPLOOP}, match={XXX}, replace={Arch. Elektr. Uebertrag. (AEU)}] %
			\step[fieldsource=\regexp{$MAPLOOP}, match={XXX}, replace={Arch. Elektron. Uebertrag. Tech.}] %
			\step[fieldsource=\regexp{$MAPLOOP}, match={XXX}, replace={Arch. Elektrotech. (Poland)}]
			\step[fieldsource=\regexp{$MAPLOOP}, match={XXX}, replace={Arch. Elektrotech. (Germany)}]
			\step[fieldsource=\regexp{$MAPLOOP}, match={XXX}, replace={Ark. Fys. Semin. Trondheim}]
			\step[fieldsource=\regexp{$MAPLOOP}, match={XXX}, replace={Astrofizika}]
			\step[fieldsource=\regexp{$MAPLOOP}, match={XXX}, replace={Astron. Nachr. (Germany)}]
			\step[fieldsource=\regexp{$MAPLOOP}, match={XXX}, replace={Astron. Tidsskr.}]
			\step[fieldsource=\regexp{$MAPLOOP}, match={XXX}, replace={Astron. Vestn.}]
			\step[fieldsource=\regexp{$MAPLOOP}, match={XXX}, replace={Astron. Zh.}]
			\step[fieldsource=\regexp{$MAPLOOP}, match={XXX}, replace={Atomwirtsch.-Atomtech.}]
			\step[fieldsource=\regexp{$MAPLOOP}, match={XXX}, replace={Atti Accad. Sci. Ist. Bologna CI. Sci. Fis.}]
			\step[fieldsource=\regexp{$MAPLOOP}, match={XXX}, replace={Rend. XIII}]
			\step[fieldsource=\regexp{$MAPLOOP}, match={XXX}, replace={Atti Accad. Sci. Torino I, CI. Sci. Fis. Math. Nat.}]
			\step[fieldsource=\regexp{$MAPLOOP}, match={XXX}, replace={Autom. Strum.}]
			\step[fieldsource=\regexp{$MAPLOOP}, match={XXX}, replace={Autom. Tech. Prax.}]
			\step[fieldsource=\regexp{$MAPLOOP}, match={XXX}, replace={Automatica}]
			\step[fieldsource=\regexp{$MAPLOOP}, match={XXX}, replace={Automatie}]
			\step[fieldsource=\regexp{$MAPLOOP}, match={XXX}, replace={Automatika}]
			\step[fieldsource=\regexp{$MAPLOOP}, match={XXX}, replace={Automatisierungstechnik}]
			\step[fieldsource=\regexp{$MAPLOOP}, match={XXX}, replace={Automatizace}]
			\step[fieldsource=\regexp{$MAPLOOP}, match={XXX}, replace={Automedica}]
			\step[fieldsource=\regexp{$MAPLOOP}, match={XXX}, replace={Avtom. Telemekh.}]
			\step[fieldsource=\regexp{$MAPLOOP}, match={XXX}, replace={Avtom. Vychisl. Tekh.}]
			\step[fieldsource=\regexp{$MAPLOOP}, match={XXX}, replace={Avtomatika}]
			\step[fieldsource=\regexp{$MAPLOOP}, match={XXX}, replace={Avtometriya}]
			\step[fieldsource=\regexp{$MAPLOOP}, match={ATZelektronik worldwide}, replace={ATZelektron. worldw.}]
			\step[fieldsource=\regexp{$MAPLOOP}, match={XXX}, replace={Ber. Bunsenges. Phys. Chem.}]
			\step[fieldsource=\regexp{$MAPLOOP}, match={XXX}, replace={Biofizika}]
			\step[fieldsource=\regexp{$MAPLOOP}, match={XXX}, replace={Biometrika}]
			\step[fieldsource=\regexp{$MAPLOOP}, match={XXX}, replace={Boll. Geofis. Teor. Appl.}]
			\step[fieldsource=\regexp{$MAPLOOP}, match={XXX}, replace={Bull. Acad. Serbe Sci. Arts cl. Sci. Tech.}]
			\step[fieldsource=\regexp{$MAPLOOP}, match={XXX}, replace={Bull. Annu. Soc. Suisse Chronom. Lab. Suisse.}]
			\step[fieldsource=\regexp{$MAPLOOP}, match={XXX}, replace={Rech. Horlog.}]
			\step[fieldsource=\regexp{$MAPLOOP}, match={XXX}, replace={Bull. Cl. Sci. Acad. R. Belg.}]
			\step[fieldsource=\regexp{$MAPLOOP}, match={XXX}, replace={Bull. Dir. Etud. Rech. A}]
			\step[fieldsource=\regexp{$MAPLOOP}, match={XXX}, replace={Bull. Dir. Etud. Rech. B}]
			\step[fieldsource=\regexp{$MAPLOOP}, match={XXX}, replace={Bull. Dir. Etud. Rech. C}]
			\step[fieldsource=\regexp{$MAPLOOP}, match={XXX}, replace={Bull. Liaison Rech. Inform. Autom.}]
			\step[fieldsource=\regexp{$MAPLOOP}, match={XXX}, replace={Bur. Etud. Autom.}]
			\step[fieldsource=\regexp{$MAPLOOP}, match={XXX}, replace={CFI-Ceram. Forum Int.-Ber. Dtsch. Keram. Ges.}]
			\step[fieldsource=\regexp{$MAPLOOP}, match={XXX}, replace={Chem. Scr.}]
			\step[fieldsource=\regexp{$MAPLOOP}, match={XXX}, replace={Ciel Terre}]
			\step[fieldsource=\regexp{$MAPLOOP}, match={XXX}, replace={Cybernetica}]
			\step[fieldsource=\regexp{$MAPLOOP}, match={XXX}, replace={Deut. Hydrogr. Z.}]
			\step[fieldsource=\regexp{$MAPLOOP}, match={XXX}, replace={Dokl. Akad. Nauk SSSR}]
			\step[fieldsource=\regexp{$MAPLOOP}, match={XXX}, replace={Electroacoustique}]
			\step[fieldsource=\regexp{$MAPLOOP}, match={XXX}, replace={Electrochim. Acta}]
			\step[fieldsource=\regexp{$MAPLOOP}, match={XXX}, replace={Electrochim. Metal.}]
			\step[fieldsource=\regexp{$MAPLOOP}, match={XXX}, replace={Elektor Electron.}]
			\step[fieldsource=\regexp{$MAPLOOP}, match={XXX}, replace={Elektr. Bahnen}]
			\step[fieldsource=\regexp{$MAPLOOP}, match={XXX}, replace={Elektr. Energ.-Tech.}]
			\step[fieldsource=\regexp{$MAPLOOP}, match={XXX}, replace={Elektr. Masch.}]
			\step[fieldsource=\regexp{$MAPLOOP}, match={XXX}, replace={Elektr. Stn.}]
			\step[fieldsource=\regexp{$MAPLOOP}, match={XXX}, replace={Elektrichestvo}]
			\step[fieldsource=\regexp{$MAPLOOP}, match={XXX}, replace={Elektrie}]
			\step[fieldsource=\regexp{$MAPLOOP}, match={XXX}, replace={Elektrizitaetswirtschaft}]
			\step[fieldsource=\regexp{$MAPLOOP}, match={XXX}, replace={Elektro}]
			\step[fieldsource=\regexp{$MAPLOOP}, match={XXX}, replace={Elektro-Anz.}]
			\step[fieldsource=\regexp{$MAPLOOP}, match={XXX}, replace={Elektro-Jahr}]
			\step[fieldsource=\regexp{$MAPLOOP}, match={XXX}, replace={Elektrokhimiya}]
			\step[fieldsource=\regexp{$MAPLOOP}, match={XXX}, replace={Elektron}]
			\step[fieldsource=\regexp{$MAPLOOP}, match={XXX}, replace={Elektron Int.}]
			\step[fieldsource=\regexp{$MAPLOOP}, match={XXX}, replace={Elektron. Entwitkl.}]
			\step[fieldsource=\regexp{$MAPLOOP}, match={XXX}, replace={Elektron. Ind}]
			\step[fieldsource=\regexp{$MAPLOOP}, match={XXX}, replace={Elektron. J.}]
			\step[fieldsource=\regexp{$MAPLOOP}, match={XXX}, replace={Elektron. Prax.}]
			\step[fieldsource=\regexp{$MAPLOOP}, match={XXX}, replace={Elektron. Tekh.}]
			\step[fieldsource=\regexp{$MAPLOOP}, match={XXX}, replace={Elektronica}]
			\step[fieldsource=\regexp{$MAPLOOP}, match={XXX}, replace={Elektronik}]
			\step[fieldsource=\regexp{$MAPLOOP}, match={XXX}, replace={Elektronika}]
			\step[fieldsource=\regexp{$MAPLOOP}, match={XXX}, replace={Elektroniker}]
			\step[fieldsource=\regexp{$MAPLOOP}, match={XXX}, replace={Elektronikschau}]
			\step[fieldsource=\regexp{$MAPLOOP}, match={XXX}, replace={Elektrosvyaz}]
			\step[fieldsource=\regexp{$MAPLOOP}, match={XXX}, replace={Elektrotech. Cas.}]
			\step[fieldsource=\regexp{$MAPLOOP}, match={Elektrotechnik und Informationstechnik}, replace={Elektrotech. Inf. Tech.}]
			\step[fieldsource=\regexp{$MAPLOOP}, match={XXX}, replace={Elektrotech. Obz.}]
			\step[fieldsource=\regexp{$MAPLOOP}, match={XXX}, replace={Elektrotechniek}]
			\step[fieldsource=\regexp{$MAPLOOP}, match={XXX}, replace={Elektrotechnik (Czechoslovakia)}]
			\step[fieldsource=\regexp{$MAPLOOP}, match={XXX}, replace={Elektrotechnik (Switzerland)}]
			\step[fieldsource=\regexp{$MAPLOOP}, match={XXX}, replace={Elektrotechnik (Germany)}]
			\step[fieldsource=\regexp{$MAPLOOP}, match={XXX}, replace={Elektrotechnika}]
			\step[fieldsource=\regexp{$MAPLOOP}, match={XXX}, replace={Elektrotehnika, Zagreb}]
			\step[fieldsource=\regexp{$MAPLOOP}, match={XXX}, replace={Elektrotekhnika}]
			\step[fieldsource=\regexp{$MAPLOOP}, match={XXX}, replace={Elektroteknikeren}]
			\step[fieldsource=\regexp{$MAPLOOP}, match={XXX}, replace={Elektrowaerme Int. B.}]
			\step[fieldsource=\regexp{$MAPLOOP}, match={XXX}, replace={Elettrificazione}]
			\step[fieldsource=\regexp{$MAPLOOP}, match={XXX}, replace={Elettron. Oggi}]
			\step[fieldsource=\regexp{$MAPLOOP}, match={XXX}, replace={Elettron. Telecomun.}]
			\step[fieldsource=\regexp{$MAPLOOP}, match={XXX}, replace={Elettrotecnica}]
			\step[fieldsource=\regexp{$MAPLOOP}, match={XXX}, replace={Elek. Med Aktuell Elektron.}]
			\step[fieldsource=\regexp{$MAPLOOP}, match={XXX}, replace={Elteknik}]
			\step[fieldsource=\regexp{$MAPLOOP}, match={XXX}, replace={Energ. Atomtech.}]
			\step[fieldsource=\regexp{$MAPLOOP}, match={XXX}, replace={Energ. Elettr.}]
			\step[fieldsource=\regexp{$MAPLOOP}, match={XXX}, replace={Energetica}]
			\step[fieldsource=\regexp{$MAPLOOP}, match={XXX}, replace={Energetik}]
			\step[fieldsource=\regexp{$MAPLOOP}, match={XXX}, replace={Energetika}]
			\step[fieldsource=\regexp{$MAPLOOP}, match={XXX}, replace={Energetyka}]
			\step[fieldsource=\regexp{$MAPLOOP}, match={XXX}, replace={Energia Nuclear}]
			\step[fieldsource=\regexp{$MAPLOOP}, match={XXX}, replace={Energie Technik (Germany)}]
			\step[fieldsource=\regexp{$MAPLOOP}, match={XXX}, replace={Energie Technik (Switzerland)}]
			\step[fieldsource=\regexp{$MAPLOOP}, match={XXX}, replace={Entropie}]
			\step[fieldsource=\regexp{$MAPLOOP}, match={XXX}, replace={ETZ}]
			\step[fieldsource=\regexp{$MAPLOOP}, match={XXX}, replace={ETZ Arch.}]
			\step[fieldsource=\regexp{$MAPLOOP}, match={XXX}, replace={Feingeraetetechnik}]
			\step[fieldsource=\regexp{$MAPLOOP}, match={XXX}, replace={Feinw. Tech. Messtech.}]
			\step[fieldsource=\regexp{$MAPLOOP}, match={XXX}, replace={Fert. Tech. Betr.}]
			\step[fieldsource=\regexp{$MAPLOOP}, match={XXX}, replace={Fis. Tecnol.}]
			\step[fieldsource=\regexp{$MAPLOOP}, match={XXX}, replace={Fiz. Khim. Obrab. Mater.}]
			\step[fieldsource=\regexp{$MAPLOOP}, match={XXX}, replace={Fiz. Met. Metalloved}]
			\step[fieldsource=\regexp{$MAPLOOP}, match={XXX}, replace={Fiz. Nizk. Temp.}]
			\step[fieldsource=\regexp{$MAPLOOP}, match={XXX}, replace={Fiz. Plazmy}]
			\step[fieldsource=\regexp{$MAPLOOP}, match={XXX}, replace={Fiz. Tekh. Poluprovodn.}]
			\step[fieldsource=\regexp{$MAPLOOP}, match={XXX}, replace={Fiz. Tverd. Tela}]
			\step[fieldsource=\regexp{$MAPLOOP}, match={XXX}, replace={Fiz.-Khim. Mekh. Mater.}]
			\step[fieldsource=\regexp{$MAPLOOP}, match={XXX}, replace={Fizika}]
			\step[fieldsource=\regexp{$MAPLOOP}, match={XXX}, replace={Forsch.-Ber. Landes Nordrh.-Westfal.}]
			\step[fieldsource=\regexp{$MAPLOOP}, match={XXX}, replace={Frequenz}]
			\step[fieldsource=\regexp{$MAPLOOP}, match={XXX}, replace={Fys. Tidsskr.}]
			\step[fieldsource=\regexp{$MAPLOOP}, match={XXX}, replace={G. Fis.}]
			\step[fieldsource=\regexp{$MAPLOOP}, match={XXX}, replace={Geliotekhnika}]
			\step[fieldsource=\regexp{$MAPLOOP}, match={XXX}, replace={Geochim. Cosmochim. Acta}]
			\step[fieldsource=\regexp{$MAPLOOP}, match={XXX}, replace={Haerterei-Tech. Mitt.}]
			\step[fieldsource=\regexp{$MAPLOOP}, match={XXX}, replace={Helv. Chim. Acta}]
			\step[fieldsource=\regexp{$MAPLOOP}, match={XXX}, replace={Helv. Med. Acta}]
			\step[fieldsource=\regexp{$MAPLOOP}, match={XXX}, replace={Helv. Phys. Acta}]
			\step[fieldsource=\regexp{$MAPLOOP}, match={XXX}, replace={Hochfreq. Electroakust}]
			\step[fieldsource=\regexp{$MAPLOOP}, match={XXX}, replace={Hoppe-Seylers Z. Physiol. Chem.}]
			\step[fieldsource=\regexp{$MAPLOOP}, match={XXX}, replace={Inf. Elektron.}]
			\step[fieldsource=\regexp{$MAPLOOP}, match={XXX}, replace={Inf. Elettron.}]
			\step[fieldsource=\regexp{$MAPLOOP}, match={XXX}, replace={Inform. Forsch. Entwickl.}]
			\step[fieldsource=\regexp{$MAPLOOP}, match={XXX}, replace={Inform. Spektrum}]
			\step[fieldsource=\regexp{$MAPLOOP}, match={XXX}, replace={Inform.-Fachber.}]
			\step[fieldsource=\regexp{$MAPLOOP}, match={XXX}, replace={Informatie}]
			\step[fieldsource=\regexp{$MAPLOOP}, match={XXX}, replace={Informatik}]
			\step[fieldsource=\regexp{$MAPLOOP}, match={XXX}, replace={Informatologia Yugosl.}]
			\step[fieldsource=\regexp{$MAPLOOP}, match={XXX}, replace={Informatyka}]
			\step[fieldsource=\regexp{$MAPLOOP}, match={XXX}, replace={Infowelt}]
			\step[fieldsource=\regexp{$MAPLOOP}, match={XXX}, replace={Ing. Electr. Mec.}]
			\step[fieldsource=\regexp{$MAPLOOP}, match={XXX}, replace={Ing. Mec. Electr.}]
			\step[fieldsource=\regexp{$MAPLOOP}, match={XXX}, replace={Ing.-Arch.}]
			\step[fieldsource=\regexp{$MAPLOOP}, match={XXX}, replace={Inzh.-Fiz. Zh.}]
			\step[fieldsource=\regexp{$MAPLOOP}, match={XXX}, replace={Izmer. Tekh.}]
			\step[fieldsource=\regexp{$MAPLOOP}, match={XXX}, replace={Izv. Akad. Nauk Arm. SSR Ser. Tekh. Nauk}]
			\step[fieldsource=\regexp{$MAPLOOP}, match={XXX}, replace={Izv. Akad. Nauk SSSR Energ. Transp.}]
			\step[fieldsource=\regexp{$MAPLOOP}, match={XXX}, replace={Izv. Akad. Nauk SSSR Fiz. Atmos. Okeana}]
			\step[fieldsource=\regexp{$MAPLOOP}, match={XXX}, replace={Izv. Akad. Nauk SSSR Fiz. Zemli}]
			\step[fieldsource=\regexp{$MAPLOOP}, match={XXX}, replace={Izv. Akad. Nauk SSSR Ser. Fiz.}]
			\step[fieldsource=\regexp{$MAPLOOP}, match={XXX}, replace={Izv. Vyssh. Uchebn. Zaved. Elektromekh.}]
			\step[fieldsource=\regexp{$MAPLOOP}, match={XXX}, replace={Izv. Vyssh. Uchebn. Zaved. Radioelektron.}]
			\step[fieldsource=\regexp{$MAPLOOP}, match={XXX}, replace={Izv. Vyssh. Uchebn. Zaved. Radiofiz.}]
			\step[fieldsource=\regexp{$MAPLOOP}, match={XXX}, replace={J. Chim Phys. Phys.-Chim Biol.}]
			\step[fieldsource=\regexp{$MAPLOOP}, match={XXX}, replace={Kernenergie}]
			\step[fieldsource=\regexp{$MAPLOOP}, match={XXX}, replace={Kerntechnik}]
			\step[fieldsource=\regexp{$MAPLOOP}, match={XXX}, replace={Khim. Fiz.}]
			\step[fieldsource=\regexp{$MAPLOOP}, match={XXX}, replace={Kibern. Vychisl. Tekh.}]
			\step[fieldsource=\regexp{$MAPLOOP}, match={XXX}, replace={Kibernetika}]
			\step[fieldsource=\regexp{$MAPLOOP}, match={XXX}, replace={Kristallografiya}]
			\step[fieldsource=\regexp{$MAPLOOP}, match={XXX}, replace={Kvantovaya Elektron. Mosk.}]
			\step[fieldsource=\regexp{$MAPLOOP}, match={XXX}, replace={Kybernetes}]
			\step[fieldsource=\regexp{$MAPLOOP}, match={XXX}, replace={Kybernetika}]
			\step[fieldsource=\regexp{$MAPLOOP}, match={XXX}, replace={Med. Tek.}]
			\step[fieldsource=\regexp{$MAPLOOP}, match={XXX}, replace={Mekh. Avtom. Proizvod.}]
			\step[fieldsource=\regexp{$MAPLOOP}, match={XXX}, replace={Meres Autom.}]
			\step[fieldsource=\regexp{$MAPLOOP}, match={XXX}, replace={Mesures}]
			\step[fieldsource=\regexp{$MAPLOOP}, match={XXX}, replace={Metallofizika}]
			\step[fieldsource=\regexp{$MAPLOOP}, match={XXX}, replace={Metalloved. Term. Obrab. Met.}]
			\step[fieldsource=\regexp{$MAPLOOP}, match={XXX}, replace={Meteorol. Gidrol.}]
			\step[fieldsource=\regexp{$MAPLOOP}, match={XXX}, replace={Meteorol. Rundsch.}]
			\step[fieldsource=\regexp{$MAPLOOP}, match={XXX}, replace={Metrol. Apl.}]
			\step[fieldsource=\regexp{$MAPLOOP}, match={XXX}, replace={Metrologia}]
			\step[fieldsource=\regexp{$MAPLOOP}, match={XXX}, replace={Medel. Simul.}]
			\step[fieldsource=\regexp{$MAPLOOP}, match={XXX}, replace={Nachr. Dok.}]
			\step[fieldsource=\regexp{$MAPLOOP}, match={XXX}, replace={Nachr.tech. Elektron.}]
			\step[fieldsource=\regexp{$MAPLOOP}, match={XXX}, replace={Naturwissentchaften}]
			\step[fieldsource=\regexp{$MAPLOOP}, match={XXX}, replace={Neue Tech.}]
			\step[fieldsource=\regexp{$MAPLOOP}, match={XXX}, replace={Neue Tech. Buero}]
			\step[fieldsource=\regexp{$MAPLOOP}, match={XXX}, replace={Nukleonika}]
			\step[fieldsource=\regexp{$MAPLOOP}, match={XXX}, replace={Numer. Math.}]
			\step[fieldsource=\regexp{$MAPLOOP}, match={XXX}, replace={Nuovo Cimento A}]
			\step[fieldsource=\regexp{$MAPLOOP}, match={XXX}, replace={Nuovo Cimento B}]
			\step[fieldsource=\regexp{$MAPLOOP}, match={XXX}, replace={Nouvo Cimento C}]
			\step[fieldsource=\regexp{$MAPLOOP}, match={XXX}, replace={Nuovo Cimento D}]
			\step[fieldsource=\regexp{$MAPLOOP}, match={XXX}, replace={Okeanologiya}]
			\step[fieldsource=\regexp{$MAPLOOP}, match={XXX}, replace={Opt. Spektrosk.}]
			\step[fieldsource=\regexp{$MAPLOOP}, match={XXX}, replace={Opt.-Mekh. Prom.}]
			\step[fieldsource=\regexp{$MAPLOOP}, match={XXX}, replace={Optik}]
			\step[fieldsource=\regexp{$MAPLOOP}, match={XXX}, replace={Photogrammetria}]
			\step[fieldsource=\regexp{$MAPLOOP}, match={XXX}, replace={Photonics Spectra}]
			\step[fieldsource=\regexp{$MAPLOOP}, match={XXX}, replace={Pis’ma Astron. Zh. }]
			\step[fieldsource=\regexp{$MAPLOOP}, match={XXX}, replace={Pis’ma Zh. Eksp. Teor. Fiz. }]
			\step[fieldsource=\regexp{$MAPLOOP}, match={XXX}, replace={Pis’ma Zh. Tekh. Fiz. }]
			\step[fieldsource=\regexp{$MAPLOOP}, match={XXX}, replace={Poverkhn., Fiz. Khim. Mekh.}]
			\step[fieldsource=\regexp{$MAPLOOP}, match={XXX}, replace={Pr. Inst. Elektrotech.}]
			\step[fieldsource=\regexp{$MAPLOOP}, match={XXX}, replace={Prib. Sist. Upr.}]
			\step[fieldsource=\regexp{$MAPLOOP}, match={XXX}, replace={Prib. Tekh. Eksp.}]
			\step[fieldsource=\regexp{$MAPLOOP}, match={XXX}, replace={Prikl. Mat. Mekh.}]
			\step[fieldsource=\regexp{$MAPLOOP}, match={XXX}, replace={Prikl. Mekh.}]
			\step[fieldsource=\regexp{$MAPLOOP}, match={XXX}, replace={Probl. Kibern.}]
			\step[fieldsource=\regexp{$MAPLOOP}, match={XXX}, replace={Probl. Peredachi Inf.}]
			\step[fieldsource=\regexp{$MAPLOOP}, match={XXX}, replace={Proc. K. Ned. Akad. Wet. B, Palaeontol.}]
			\step[fieldsource=\regexp{$MAPLOOP}, match={XXX}, replace={Anthropol. }]
			\step[fieldsource=\regexp{$MAPLOOP}, match={XXX}, replace={Programmirovanie}]
			\step[fieldsource=\regexp{$MAPLOOP}, match={XXX}, replace={Prz. Elektrotech.}]
			\step[fieldsource=\regexp{$MAPLOOP}, match={XXX}, replace={Prz. Telekomun.}]
			\step[fieldsource=\regexp{$MAPLOOP}, match={XXX}, replace={PT/Elektrotech. Elektron.}]
			\step[fieldsource=\regexp{$MAPLOOP}, match={XXX}, replace={Radio Fernsehen Elektron.}]
			\step[fieldsource=\regexp{$MAPLOOP}, match={XXX}, replace={Radiotekh. Elektron.}]
			\step[fieldsource=\regexp{$MAPLOOP}, match={XXX}, replace={Radiotekhnika Mosk.}]
			\step[fieldsource=\regexp{$MAPLOOP}, match={XXX}, replace={Rev. Acad. Cienc. Zaragoza}]
			\step[fieldsource=\regexp{$MAPLOOP}, match={XXX}, replace={Rev. Electrotec. (Argentina)}]
			\step[fieldsource=\regexp{$MAPLOOP}, match={XXX}, replace={Rev. Electrotec. (Spain)}]
			\step[fieldsource=\regexp{$MAPLOOP}, match={XXX}, replace={Rev. Energ.}]
			\step[fieldsource=\regexp{$MAPLOOP}, match={XXX}, replace={Rev. Esp. Electron. Rev. Geofis.}]
			\step[fieldsource=\regexp{$MAPLOOP}, match={XXX}, replace={Ric. Autom.}]
			\step[fieldsource=\regexp{$MAPLOOP}, match={XXX}, replace={Ric. Spettrosc.}]
			\step[fieldsource=\regexp{$MAPLOOP}, match={XXX}, replace={Robotersysteme}]
			\step[fieldsource=\regexp{$MAPLOOP}, match={XXX}, replace={Rozpr. Electrotech.}]
			\step[fieldsource=\regexp{$MAPLOOP}, match={XXX}, replace={Sadhana}]
			\step[fieldsource=\regexp{$MAPLOOP}, match={XXX}, replace={Schweiz. Tech. Z.}]
			\step[fieldsource=\regexp{$MAPLOOP}, match={XXX}, replace={Scientia}]
			\step[fieldsource=\regexp{$MAPLOOP}, match={XXX}, replace={Siemens Forsch. Entwickl. Ber.}]
			\step[fieldsource=\regexp{$MAPLOOP}, match={XXX}, replace={Sist. Autom.}]
			\step[fieldsource=\regexp{$MAPLOOP}, match={XXX}, replace={Sitzungsber. Oester. Akad. Wiss. Math.-}]
			\step[fieldsource=\regexp{$MAPLOOP}, match={XXX}, replace={Naturwiss. Kl. Abt. II (Austria)}]
			\step[fieldsource=\regexp{$MAPLOOP}, match={XXX}, replace={Spectrochim. Acta A, Mol. Spectrosc.}]
			\step[fieldsource=\regexp{$MAPLOOP}, match={XXX}, replace={Spectrochim. Acta B, At. Spectrosc.}]
			\step[fieldsource=\regexp{$MAPLOOP}, match={XXX}, replace={Sprache Datenverarb.}]
			\step[fieldsource=\regexp{$MAPLOOP}, match={XXX}, replace={Stanki Instrum.}]
			\step[fieldsource=\regexp{$MAPLOOP}, match={XXX}, replace={Steklo Keram.}]
			\step[fieldsource=\regexp{$MAPLOOP}, match={XXX}, replace={Svetotekhnika  TE Int.}]
			\step[fieldsource=\regexp{$MAPLOOP}, match={XXX}, replace={Tech. Bull. Vevey}]
			\step[fieldsource=\regexp{$MAPLOOP}, match={XXX}, replace={Tech. Mitt. Krupp (Engl. Ed.)}]
			\step[fieldsource=\regexp{$MAPLOOP}, match={XXX}, replace={Tech. Mitt. PTT}]
			\step[fieldsource=\regexp{$MAPLOOP}, match={XXX}, replace={Tech. Mitt. RFZ}]
			\step[fieldsource=\regexp{$MAPLOOP}, match={XXX}, replace={Technica}]
			\step[fieldsource=\regexp{$MAPLOOP}, match={XXX}, replace={Tecnica}]
			\step[fieldsource=\regexp{$MAPLOOP}, match={XXX}, replace={Teh. Fiz.}]
			\step[fieldsource=\regexp{$MAPLOOP}, match={XXX}, replace={Tehnika}]
			\step[fieldsource=\regexp{$MAPLOOP}, match={XXX}, replace={Tekh. Elektrodin.}]
			\step[fieldsource=\regexp{$MAPLOOP}, match={XXX}, replace={Tekh. Kibern.}]
			\step[fieldsource=\regexp{$MAPLOOP}, match={XXX}, replace={Tekh. Kino Telev.}]
			\step[fieldsource=\regexp{$MAPLOOP}, match={XXX}, replace={Tekh. Misul}]
			\step[fieldsource=\regexp{$MAPLOOP}, match={XXX}, replace={Telekomunikacije}]
			\step[fieldsource=\regexp{$MAPLOOP}, match={XXX}, replace={Telektronikk}]
			\step[fieldsource=\regexp{$MAPLOOP}, match={XXX}, replace={Teleteknik}]
			\step[fieldsource=\regexp{$MAPLOOP}, match={XXX}, replace={Teor. Mat. Fiz.}]
			\step[fieldsource=\regexp{$MAPLOOP}, match={XXX}, replace={Teploeoergetika}]
			\step[fieldsource=\regexp{$MAPLOOP}, match={XXX}, replace={Teplofiz. Vvs. Temp.}]
			\step[fieldsource=\regexp{$MAPLOOP}, match={XXX}, replace={Tidskr. Dok.}]
			\step[fieldsource=\regexp{$MAPLOOP}, match={XXX}, replace={TN Nachr.}]
			\step[fieldsource=\regexp{$MAPLOOP}, match={XXX}, replace={Toute Electron.}]
			\step[fieldsource=\regexp{$MAPLOOP}, match={XXX}, replace={Tr. Inst. Teor. Astron.}]
			\step[fieldsource=\regexp{$MAPLOOP}, match={XXX}, replace={Ukr. Fiz. Zh.}]
			\step[fieldsource=\regexp{$MAPLOOP}, match={XXX}, replace={Usp. Fiz. Nauk}]
			\step[fieldsource=\regexp{$MAPLOOP}, match={XXX}, replace={Vak.-Tech.}]
			\step[fieldsource=\regexp{$MAPLOOP}, match={XXX}, replace={VDE Fachiber.}]
			\step[fieldsource=\regexp{$MAPLOOP}, match={XXX}, replace={VDI Z.}]
			\step[fieldsource=\regexp{$MAPLOOP}, match={XXX}, replace={Vestn. Mashinostr.}]
			\step[fieldsource=\regexp{$MAPLOOP}, match={XXX}, replace={Vestn. Mosk. Univ. 15, Vychisl. Mat. Kibern.}]
			\step[fieldsource=\regexp{$MAPLOOP}, match={XXX}, replace={Vestn. Mosk. Univ. 3, Fiz. Astron.}]
			\step[fieldsource=\regexp{$MAPLOOP}, match={XXX}, replace={Vesti Akad. Navuk BSSR Ser. Fiz. Energ. Navuk}]
			\step[fieldsource=\regexp{$MAPLOOP}, match={XXX}, replace={VGB Kraftwerkstech. (Ger. Ed.)}]
			\step[fieldsource=\regexp{$MAPLOOP}, match={XXX}, replace={Vide Couches Minces}]
			\step[fieldsource=\regexp{$MAPLOOP}, match={XXX}, replace={Vistas Astron.}]
			\step[fieldsource=\regexp{$MAPLOOP}, match={XXX}, replace={Vopr. At. Nauki Tekh. Ser., Fiz. Radiats.}]
			\step[fieldsource=\regexp{$MAPLOOP}, match={XXX}, replace={Povrezhdenii Radiats. Materialoved.}]
			\step[fieldsource=\regexp{$MAPLOOP}, match={XXX}, replace={Vopr. At. Nauki Tekh. Ser., Obshch. Yad. Fiz.}]
			\step[fieldsource=\regexp{$MAPLOOP}, match={XXX}, replace={Vuoto Sci. Tecnol.}]
			\step[fieldsource=\regexp{$MAPLOOP}, match={XXX}, replace={Wiss. Z. Friedrich-Schiller-Univ. Jena}]
			\step[fieldsource=\regexp{$MAPLOOP}, match={XXX}, replace={Nat.wiss. Reihe}]
			\step[fieldsource=\regexp{$MAPLOOP}, match={XXX}, replace={Wiss. Z. Karl-Marx-Univ. Leipz. Math.-}]
			\step[fieldsource=\regexp{$MAPLOOP}, match={XXX}, replace={Nat.wiss. Reihe}]
			\step[fieldsource=\regexp{$MAPLOOP}, match={XXX}, replace={Wiss. Z. Tech. Hochsch. Ilmenau}]
			\step[fieldsource=\regexp{$MAPLOOP}, match={XXX}, replace={Wiss. Z. Tech. Univ. Dresd.}]
			\step[fieldsource=\regexp{$MAPLOOP}, match={XXX}, replace={Wiss. Z. Tech. Univ. Karl-Marx-Stadt}]
			\step[fieldsource=\regexp{$MAPLOOP}, match={XXX}, replace={Yad. Fiz.}]
			\step[fieldsource=\regexp{$MAPLOOP}, match={XXX}, replace={Z. Angew. Math. Mech.}]
			\step[fieldsource=\regexp{$MAPLOOP}, match={XXX}, replace={Z. Angew. Math. Phys.}]
			\step[fieldsource=\regexp{$MAPLOOP}, match={XXX}, replace={Z. Met.kd.}]
			\step[fieldsource=\regexp{$MAPLOOP}, match={XXX}, replace={Z. Nat. Forsch. A, Phys. Phys. Chem. Kosmophys.}]
			\step[fieldsource=\regexp{$MAPLOOP}, match={XXX}, replace={Z. Oper. Res. A, Theor.}]
			\step[fieldsource=\regexp{$MAPLOOP}, match={XXX}, replace={Z. Oper. Res. B, Prax.}]
			\step[fieldsource=\regexp{$MAPLOOP}, match={XXX}, replace={Z. Phys.}]
			\step[fieldsource=\regexp{$MAPLOOP}, match={XXX}, replace={Z. Phys. A, At. Nuclei}]
			\step[fieldsource=\regexp{$MAPLOOP}, match={XXX}, replace={Z. Phys. B, Condens. Matter}]
			\step[fieldsource=\regexp{$MAPLOOP}, match={XXX}, replace={Z. Phys. C, Part Fields}]
			\step[fieldsource=\regexp{$MAPLOOP}, match={XXX}, replace={Z. Phys. Chem. Neue Folge}]
			\step[fieldsource=\regexp{$MAPLOOP}, match={XXX}, replace={Z. Phys. Chem., Leipz.}]
			\step[fieldsource=\regexp{$MAPLOOP}, match={XXX}, replace={Z. Phys. D, At. Mol. Clusters}]
			\step[fieldsource=\regexp{$MAPLOOP}, match={XXX}, replace={Zavod. Lab.}]
			\step[fieldsource=\regexp{$MAPLOOP}, match={XXX}, replace={Zh. Eksp. Teor. Fiz.}]
			\step[fieldsource=\regexp{$MAPLOOP}, match={XXX}, replace={Zh. Fiz. Khim.}]
			\step[fieldsource=\regexp{$MAPLOOP}, match={XXX}, replace={Zh. Prikl. Mekh. Tekh. Fiz.}]
			\step[fieldsource=\regexp{$MAPLOOP}, match={XXX}, replace={Zh. Prikl. Spektrosk.}]
			\step[fieldsource=\regexp{$MAPLOOP}, match={XXX}, replace={Zh. Tekh. Fiz.}]
			\step[fieldsource=\regexp{$MAPLOOP}, match={XXX}, replace={Zh. Vychisl. Mat. Mat. Fiz.}]
			\step[fieldsource=\regexp{$MAPLOOP}, match={XXX}, replace={Zisin, J. Seismol. Soc. Jpn.}] 
			\step[fieldsource=\regexp{$MAPLOOP}, match={Production}, replace={Prod.}]
			\step[fieldsource=\regexp{$MAPLOOP}, match={Reliability}, replace={Rel.}]
			\step[fieldsource=\regexp{$MAPLOOP}, match={Report}, replace={Rep.}]
			\step[fieldsource=\regexp{$MAPLOOP}, match={Semiconductor}, replace={Semicond.}]
			\step[fieldsource=\regexp{$MAPLOOP}, match={Research}, replace={Res.}]
			\step[fieldsource=\regexp{$MAPLOOP}, match={Sensing}, replace={Sens.}]
			\step[fieldsource=\regexp{$MAPLOOP}, match={Resonance}, replace={Reson.}]
			\step[fieldsource=\regexp{$MAPLOOP}, match={Series}, replace={Ser.}]
			\step[fieldsource=\regexp{$MAPLOOP}, match={Resources}, replace={Resour.}]
			\step[fieldsource=\regexp{$MAPLOOP}, match={Simulation}, replace={Simul.}]
			\step[fieldsource=\regexp{$MAPLOOP}, match={Reviews}, replace={Rev.}]
			\step[fieldsource=\regexp{$MAPLOOP}, match={Review}, replace={Rev.}]
			\step[fieldsource=\regexp{$MAPLOOP}, match={Singapore}, replace={Singap.}]
			\step[fieldsource=\regexp{$MAPLOOP}, match={Robotics}, replace={Robot.}]
			\step[fieldsource=\regexp{$MAPLOOP}, match={Sistema}, replace={Sist.}]
			\step[fieldsource=\regexp{$MAPLOOP}, match={Royal}, replace={Roy.}]
			\step[fieldsource=\regexp{$MAPLOOP}, match={Society}, replace={Soc.}]
			\step[fieldsource=\regexp{$MAPLOOP}, match={Safety}, replace={Saf.}]
			\step[fieldsource=\regexp{$MAPLOOP}, match={Sociological}, replace={Sociol.}]
			\step[fieldsource=\regexp{$MAPLOOP}, match={Satellite}, replace={Satell.}]
			\step[fieldsource=\regexp{$MAPLOOP}, match={Software}, replace={Softw.}]
			\step[fieldsource=\regexp{$MAPLOOP}, match={Scandinavian}, replace={Scand.}]
			\step[fieldsource=\regexp{$MAPLOOP}, match={Solar}, replace={Sol.}]
			\step[fieldsource=\regexp{$MAPLOOP}, match={Sciences}, replace={Sci.}]
			\step[fieldsource=\regexp{$MAPLOOP}, match={Science}, replace={Sci.}]
			\step[fieldsource=\regexp{$MAPLOOP}, match={Soviet}, replace={Sov.}]
			\step[fieldsource=\regexp{$MAPLOOP}, match={Section}, replace={Sect.}]
			\step[fieldsource=\regexp{$MAPLOOP}, match={Spectroscopy}, replace={Spectrosc.}]
			\step[fieldsource=\regexp{$MAPLOOP}, match={Security}, replace={Secur.}]
			\step[fieldsource=\regexp{$MAPLOOP}, match={Spectrum}, replace={Spectr.}]
			\step[fieldsource=\regexp{$MAPLOOP}, match={Seismology}, replace={Seismol.}]
			\step[fieldsource=\regexp{$MAPLOOP}, match={Speculations}, replace={Specul.}]
			\step[fieldsource=\regexp{$MAPLOOP}, match={Selected}, replace={Sel.}]
			\step[fieldsource=\regexp{$MAPLOOP}, match={Statistics}, replace={Statist.}]
			\step[fieldsource=\regexp{$MAPLOOP}, match={Structures}, replace={Struct.}]
			\step[fieldsource=\regexp{$MAPLOOP}, match={Structure}, replace={Struct.}]
			\step[fieldsource=\regexp{$MAPLOOP}, match={Terrestrial}, replace={Terr.}]
			\step[fieldsource=\regexp{$MAPLOOP}, match={Studies}, replace={Stud.}]
			\step[fieldsource=\regexp{$MAPLOOP}, match={Theoretical}, replace={Theor.}]
			\step[fieldsource=\regexp{$MAPLOOP}, match={Superconductivity}, replace={Supercond.}]
			\step[fieldsource=\regexp{$MAPLOOP}, match={Transactions}, replace={Trans.}]
			\step[fieldsource=\regexp{$MAPLOOP}, match={Supplement}, replace={Suppl.}]
			\step[fieldsource=\regexp{$MAPLOOP}, match={Translation}, replace={Transl.}]
			\step[fieldsource=\regexp{$MAPLOOP}, match={Surface}, replace={Surf.}]
			\step[fieldsource=\regexp{$MAPLOOP}, match={Transmission}, replace={Transmiss.}]
			\step[fieldsource=\regexp{$MAPLOOP}, match={Survey}, replace={Surv.}]
			\step[fieldsource=\regexp{$MAPLOOP}, match={Transportation}, replace={Transp.}]
			\step[fieldsource=\regexp{$MAPLOOP}, match={Sustainable}, replace={Sustain.}]
			\step[fieldsource=\regexp{$MAPLOOP}, match={Tutorials}, replace={Tut.}]
			\step[fieldsource=\regexp{$MAPLOOP}, match={Symposium}, replace={Symp.}]
			\step[fieldsource=\regexp{$MAPLOOP}, match={Ultrasonic}, replace={Ultrason.}]
			\step[fieldsource=\regexp{$MAPLOOP}, match={Systems}, replace={Syst.}]
			\step[fieldsource=\regexp{$MAPLOOP}, match={System}, replace={Syst.}]
			\step[fieldsource=\regexp{$MAPLOOP}, match={University}, replace={Univ.}]
			\step[fieldsource=\regexp{$MAPLOOP}, match={\detokenize{Universität}}, replace={Univ.}]
			\step[fieldsource=\regexp{$MAPLOOP}, match={\detokenize{Université}}, replace={Univ.}]
			\step[fieldsource=\regexp{$MAPLOOP}, match={Technical}, replace={Tech.}]
			\step[fieldsource=\regexp{$MAPLOOP}, match={Technische}, replace={Tech.}]
			\step[fieldsource=\regexp{$MAPLOOP}, match={Vacuum}, replace={Vac.}]
			\step[fieldsource=\regexp{$MAPLOOP}, match={Techniques}, replace={Techn.}]
			\step[fieldsource=\regexp{$MAPLOOP}, match={Vehicles}, replace={Veh.}]
			\step[fieldsource=\regexp{$MAPLOOP}, match={Vehicle}, replace={Veh.}]
			\step[fieldsource=\regexp{$MAPLOOP}, match={Vehicular}, replace={Veh.}]
			\step[fieldsource=\regexp{$MAPLOOP}, match={Technology}, replace={Technol.}]
			\step[fieldsource=\regexp{$MAPLOOP}, match={Technological}, replace={Technol.}]
			\step[fieldsource=\regexp{$MAPLOOP}, match={Vibration}, replace={Vib.}]
			\step[fieldsource=\regexp{$MAPLOOP}, match={Telecommunications}, replace={Telecommun.}]
			\step[fieldsource=\regexp{$MAPLOOP}, match={Visual}, replace={Vis.}]
			\step[fieldsource=\regexp{$MAPLOOP}, match={Television}, replace={Telev.}]
			\step[fieldsource=\regexp{$MAPLOOP}, match={Welding}, replace={Weld.}]
			\step[fieldsource=\regexp{$MAPLOOP}, match={Temperature}, replace={Temp.}]
			\step[fieldsource=\regexp{$MAPLOOP}, match={Working}, replace={Work.}]
			\step[fieldsource=\regexp{$MAPLOOP}, match={Learning}, replace={Learn.}]
			\step[fieldsource=\regexp{$MAPLOOP}, match={Measurement}, replace={Meas.}]
			\step[fieldsource=\regexp{$MAPLOOP}, match={Letters}, replace={Lett.}]
			\step[fieldsource=\regexp{$MAPLOOP}, match={Letter}, replace={Lett.}]
			\step[fieldsource=\regexp{$MAPLOOP}, match={Mechanical}, replace={Mech.}]
			\step[fieldsource=\regexp{$MAPLOOP}, match={Mechanics}, replace={Mech.}]
			\step[fieldsource=\regexp{$MAPLOOP}, match={Mechanic}, replace={Mech.}]
			\step[fieldsource=\regexp{$MAPLOOP}, match={Lightwave}, replace={Lightw.}]
			\step[fieldsource=\regexp{$MAPLOOP}, match={Medical}, replace={Med.}]
			\step[fieldsource=\regexp{$MAPLOOP}, match={Logic, Logical}, replace={Log.}]
			\step[fieldsource=\regexp{$MAPLOOP}, match={Metals}, replace={Met.}]
			\step[fieldsource=\regexp{$MAPLOOP}, match={Luminescence}, replace={Lumin.}]
			\step[fieldsource=\regexp{$MAPLOOP}, match={Metallurgy}, replace={Metall.}]
			\step[fieldsource=\regexp{$MAPLOOP}, match={Machines}, replace={Mach.}]
			\step[fieldsource=\regexp{$MAPLOOP}, match={Machine}, replace={Mach.}]
			\step[fieldsource=\regexp{$MAPLOOP}, match={Meteorology}, replace={Meteorol.}]
			\step[fieldsource=\regexp{$MAPLOOP}, match={Magazine}, replace={Mag.}]
			\step[fieldsource=\regexp{$MAPLOOP}, match={Metropolitan}, replace={Metrop.}]
			\step[fieldsource=\regexp{$MAPLOOP}, match={Magnetics}, replace={Magn.}]
			\step[fieldsource=\regexp{$MAPLOOP}, match={Mexican, Mexico}, replace={Mex.}]
			\step[fieldsource=\regexp{$MAPLOOP}, match={Management}, replace={Manage.}]
			\step[fieldsource=\regexp{$MAPLOOP}, match={Microelectromechanical}, replace={Microelectromech.}]
			\step[fieldsource=\regexp{$MAPLOOP}, match={Managing}, replace={Manag.}]
			\step[fieldsource=\regexp{$MAPLOOP}, match={Microgravity}, replace={Microgr.}]
			\step[fieldsource=\regexp{$MAPLOOP}, match={Manufacturing}, replace={Manuf.}]
			\step[fieldsource=\regexp{$MAPLOOP}, match={Microscopy}, replace={Microsc.}]
			\step[fieldsource=\regexp{$MAPLOOP}, match={Marine}, replace={Mar.}]
			\step[fieldsource=\regexp{$MAPLOOP}, match={Microwaves}, replace={Microw.}]
			\step[fieldsource=\regexp{$MAPLOOP}, match={Microwave}, replace={Microw.}]
			\step[fieldsource=\regexp{$MAPLOOP}, match={Materials}, replace={Mater.}]
			\step[fieldsource=\regexp{$MAPLOOP}, match={Material}, replace={Mater.}]
			\step[fieldsource=\regexp{$MAPLOOP}, match={Military}, replace={Mil.}]
			\step[fieldsource=\regexp{$MAPLOOP}, match={Modeling}, replace={Model.}]
			\step[fieldsource=\regexp{$MAPLOOP}, match={Modelling}, replace={Model.}]
			\step[fieldsource=\regexp{$MAPLOOP}, match={Oceanic}, replace={Ocean.}]
			\step[fieldsource=\regexp{$MAPLOOP}, match={Molecular}, replace={Mol.}]
			\step[fieldsource=\regexp{$MAPLOOP}, match={Oceanography}, replace={Oceanogr.}]
			\step[fieldsource=\regexp{$MAPLOOP}, match={Monitoring}, replace={Monit.}]
			\step[fieldsource=\regexp{$MAPLOOP}, match={Occupation}, replace={Occupat.}]
			\step[fieldsource=\regexp{$MAPLOOP}, match={Multiphysics}, replace={Multiphys.}]
			\step[fieldsource=\regexp{$MAPLOOP}, match={Operational}, replace={Oper.}]
			\step[fieldsource=\regexp{$MAPLOOP}, match={Nanobioscience}, replace={Nanobiosci.}]
			\step[fieldsource=\regexp{$MAPLOOP}, match={Optical}, replace={Opt.}]
			\step[fieldsource=\regexp{$MAPLOOP}, match={Nanotechnology}, replace={Nanotechnol.}]
			\step[fieldsource=\regexp{$MAPLOOP}, match={Optics}, replace={Opt.}]
			\step[fieldsource=\regexp{$MAPLOOP}, match={National}, replace={Nat.}]
			\step[fieldsource=\regexp{$MAPLOOP}, match={Optimization}, replace={Optim.}]
			\step[fieldsource=\regexp{$MAPLOOP}, match={Naval}, replace={Nav.}]
			\step[fieldsource=\regexp{$MAPLOOP}, match={Organization}, replace={Org.}]
			\step[fieldsource=\regexp{$MAPLOOP}, match={Networking}, replace={Netw.}]
			\step[fieldsource=\regexp{$MAPLOOP}, match={Networked}, replace={Netw.}]
			\step[fieldsource=\regexp{$MAPLOOP}, match={Network}, replace={Netw.}]
			\step[fieldsource=\regexp{$MAPLOOP}, match={Packaging}, replace={Packag.}]
			\step[fieldsource=\regexp{$MAPLOOP}, match={Newsletter}, replace={Newslett.}]
			\step[fieldsource=\regexp{$MAPLOOP}, match={Particle}, replace={Part.}]
			\step[fieldsource=\regexp{$MAPLOOP}, match={Nondestructive}, replace={Nondestruct.}]
			\step[fieldsource=\regexp{$MAPLOOP}, match={Patent}, replace={Pat.}]
			\step[fieldsource=\regexp{$MAPLOOP}, match={Nuclear}, replace={Nucl.}]
			\step[fieldsource=\regexp{$MAPLOOP}, match={Performance}, replace={Perform.}]
			\step[fieldsource=\regexp{$MAPLOOP}, match={Numerical}, replace={Numer.}]
			\step[fieldsource=\regexp{$MAPLOOP}, match={Personal}, replace={Pers.}]
			\step[fieldsource=\regexp{$MAPLOOP}, match={Observations}, replace={Observ.}]
			\step[fieldsource=\regexp{$MAPLOOP}, match={Philosophical}, replace={Philos.}]
			\step[fieldsource=\regexp{$MAPLOOP}, match={Photonics}, replace={Photon.}]
			\step[fieldsource=\regexp{$MAPLOOP}, match={Productivity}, replace={Productiv.}]
			\step[fieldsource=\regexp{$MAPLOOP}, match={Photovoltaics}, replace={Photovolt.}]
			\step[fieldsource=\regexp{$MAPLOOP}, match={Programming}, replace={Program.}]
			\step[fieldsource=\regexp{$MAPLOOP}, match={Physics}, replace={Phys.}]
			\step[fieldsource=\regexp{$MAPLOOP}, match={Progress}, replace={Prog.}]
			\step[fieldsource=\regexp{$MAPLOOP}, match={Physiology}, replace={Physiol.}]
			\step[fieldsource=\regexp{$MAPLOOP}, match={Propagation}, replace={Propag.}]
			\step[fieldsource=\regexp{$MAPLOOP}, match={Planetary}, replace={Planet.}]
			\step[fieldsource=\regexp{$MAPLOOP}, match={Psychology}, replace={Psychol.}]
			\step[fieldsource=\regexp{$MAPLOOP}, match={Pneumatics}, replace={Pneum.}]
			\step[fieldsource=\regexp{$MAPLOOP}, match={Quality}, replace={Qual.}]
			\step[fieldsource=\regexp{$MAPLOOP}, match={Pollution}, replace={Pollut.}]
			\step[fieldsource=\regexp{$MAPLOOP}, match={Quarterly}, replace={Quart.}]
			\step[fieldsource=\regexp{$MAPLOOP}, match={Polymer}, replace={Polym.}]
			\step[fieldsource=\regexp{$MAPLOOP}, match={Radiation}, replace={Radiat.}]
			\step[fieldsource=\regexp{$MAPLOOP}, match={Polytechnic}, replace={Polytech.}]
			\step[fieldsource=\regexp{$MAPLOOP}, match={Radiology}, replace={Radiol.}]
			\step[fieldsource=\regexp{$MAPLOOP}, match={Practice}, replace={Pract.}]
			\step[fieldsource=\regexp{$MAPLOOP}, match={Reactor}, replace={React.}]
			\step[fieldsource=\regexp{$MAPLOOP}, match={Precision}, replace={Precis.}]
			\step[fieldsource=\regexp{$MAPLOOP}, match={Receivers}, replace={Receiv.}]
			\step[fieldsource=\regexp{$MAPLOOP}, match={Principles}, replace={Princ.}]
			\step[fieldsource=\regexp{$MAPLOOP}, match={Recognition}, replace={Recognit.}]
			\step[fieldsource=\regexp{$MAPLOOP}, match={Proceedings}, replace={Proc.}]
			\step[fieldsource=\regexp{$MAPLOOP}, match={Record}, replace={Rec.}]
			\step[fieldsource=\regexp{$MAPLOOP}, match={Processing}, replace={Process.}]
			\step[fieldsource=\regexp{$MAPLOOP}, match={Rehabilitation}, replace={Rehabil.}]
			\step[fieldsource=\regexp{$MAPLOOP}, match={Conversion}, replace={Convers.}]
			\step[fieldsource=\regexp{$MAPLOOP}, match={Digital}, replace={Digit.}]
			\step[fieldsource=\regexp{$MAPLOOP}, match={Convention}, replace={Conv.}]
			\step[fieldsource=\regexp{$MAPLOOP}, match={Disclosure}, replace={Discl.}]
			\step[fieldsource=\regexp{$MAPLOOP}, match={Correspondence}, replace={Corresp.}]
			\step[fieldsource=\regexp{$MAPLOOP}, match={Discussions}, replace={Discuss.}]
			\step[fieldsource=\regexp{$MAPLOOP}, match={Critical}, replace={Crit.}]
			\step[fieldsource=\regexp{$MAPLOOP}, match={Dissertations}, replace={Diss.}]
			\step[fieldsource=\regexp{$MAPLOOP}, match={Crystal}, replace={Cryst.}]
			\step[fieldsource=\regexp{$MAPLOOP}, match={Distributed}, replace={Distrib.}]
			\step[fieldsource=\regexp{$MAPLOOP}, match={Crystallography}, replace={Crystallogr.}]
			\step[fieldsource=\regexp{$MAPLOOP}, match={Dynamics}, replace={Dyn.}]
			\step[fieldsource=\regexp{$MAPLOOP}, match={Cybernetics}, replace={Cybern.}]
			\step[fieldsource=\regexp{$MAPLOOP}, match={Earthquake}, replace={Earthq.}]
			\step[fieldsource=\regexp{$MAPLOOP}, match={Decision}, replace={Decis.}]
			\step[fieldsource=\regexp{$MAPLOOP}, match={Economics}, replace={Econ.}]
			\step[fieldsource=\regexp{$MAPLOOP}, match={Economic}, replace={Econ.}]
			\step[fieldsource=\regexp{$MAPLOOP}, match={Economical}, replace={Econ.}]
			\step[fieldsource=\regexp{$MAPLOOP}, match={Edition}, replace={Ed.}]
			\step[fieldsource=\regexp{$MAPLOOP}, match={Evolutionary}, replace={Evol.}]
			\step[fieldsource=\regexp{$MAPLOOP}, match={Education}, replace={Educ.}]
			\step[fieldsource=\regexp{$MAPLOOP}, match={Exhibition}, replace={Exhib.}]
			\step[fieldsource=\regexp{$MAPLOOP}, match={Electrical}, replace={Elect.}]
			\step[fieldsource=\regexp{$MAPLOOP}, match={Electric}, replace={Elect.}]
			\step[fieldsource=\regexp{$MAPLOOP}, match={Experimental}, replace={Exp.}]
			\step[fieldsource=\regexp{$MAPLOOP}, match={Electrification}, replace={Electrific.}]
			\step[fieldsource=\regexp{$MAPLOOP}, match={Exploratory}, replace={Explor.}]
			\step[fieldsource=\regexp{$MAPLOOP}, match={Electromagnetic}, replace={Electromagn.}]
			\step[fieldsource=\regexp{$MAPLOOP}, match={Exposition}, replace={Expo.}]
			\step[fieldsource=\regexp{$MAPLOOP}, match={Electroacoustic}, replace={Electroacoust.}]
			\step[fieldsource=\regexp{$MAPLOOP}, match={Express}, replace={Express}]
			\step[fieldsource=\regexp{$MAPLOOP}, match={Electronics}, replace={Electron.}]
			\step[fieldsource=\regexp{$MAPLOOP}, match={Electronic}, replace={Electron.}]
			\step[fieldsource=\regexp{$MAPLOOP}, match={Fabrication}, replace={Fabr.}]
			\step[fieldsource=\regexp{$MAPLOOP}, match={Emerging}, replace={Emerg.}]
			\step[fieldsource=\regexp{$MAPLOOP}, match={Faculty}, replace={Fac.}]
			\step[fieldsource=\regexp{$MAPLOOP}, match={Engineering}, replace={Eng.}]
			\step[fieldsource=\regexp{$MAPLOOP}, match={Engineers}, replace={Eng.}]
			\step[fieldsource=\regexp{$MAPLOOP}, match={Engineer}, replace={Eng.}]
			\step[fieldsource=\regexp{$MAPLOOP}, match={Ferroelectrics}, replace={Ferroelect.}]
			\step[fieldsource=\regexp{$MAPLOOP}, match={Environment}, replace={Environ.}]
			\step[fieldsource=\regexp{$MAPLOOP}, match={Francais, French}, replace={Fr.}]
			\step[fieldsource=\regexp{$MAPLOOP}, match={Equations}, replace={Equ.}]
			\step[fieldsource=\regexp{$MAPLOOP}, match={Frequency}, replace={Freq.}]
			\step[fieldsource=\regexp{$MAPLOOP}, match={Equipment}, replace={Equip.}]
			\step[fieldsource=\regexp{$MAPLOOP}, match={Foundation}, replace={Found.}]
			\step[fieldsource=\regexp{$MAPLOOP}, match={Ergonomics}, replace={Ergonom.}]
			\step[fieldsource=\regexp{$MAPLOOP}, match={Fundamental}, replace={Fundam.}]
			\step[fieldsource=\regexp{$MAPLOOP}, match={European}, replace={Eur.}]
			\step[fieldsource=\regexp{$MAPLOOP}, match={Generation}, replace={Gener.}]
			\step[fieldsource=\regexp{$MAPLOOP}, match={Evaluation}, replace={Eval.}]
			\step[fieldsource=\regexp{$MAPLOOP}, match={Geology}, replace={Geol.}]
			\step[fieldsource=\regexp{$MAPLOOP}, match={Geophysics}, replace={Geophys.}]
			\step[fieldsource=\regexp{$MAPLOOP}, match={Innovations}, replace={Innov.}]
			\step[fieldsource=\regexp{$MAPLOOP}, match={Innovation}, replace={Innov.}]
			\step[fieldsource=\regexp{$MAPLOOP}, match={Geoscience}, replace={Geosci.}]
			\step[fieldsource=\regexp{$MAPLOOP}, match={Institute}, replace={Inst.}]
			\step[fieldsource=\regexp{$MAPLOOP}, match={Institution}, replace={Inst.}]
			\step[fieldsource=\regexp{$MAPLOOP}, match={Graphics}, replace={Graph.}]
			\step[fieldsource=\regexp{$MAPLOOP}, match={Instrument}, replace={Instrum.}]
			\step[fieldsource=\regexp{$MAPLOOP}, match={Guidance}, replace={Guid.}]
			\step[fieldsource=\regexp{$MAPLOOP}, match={Instrumentation}, replace={Instrum.}]
			\step[fieldsource=\regexp{$MAPLOOP}, match={Harmonics}, replace={Harmon.}]
			\step[fieldsource=\regexp{$MAPLOOP}, match={Harmonic}, replace={Harmon.}]
			\step[fieldsource=\regexp{$MAPLOOP}, match={Insulation}, replace={Insul.}]
			\step[fieldsource=\regexp{$MAPLOOP}, match={History}, replace={Hist.}]
			\step[fieldsource=\regexp{$MAPLOOP}, match={Integrated}, replace={Integr.}]
			\step[fieldsource=\regexp{$MAPLOOP}, match={Horizon}, replace={Horiz.}]
			\step[fieldsource=\regexp{$MAPLOOP}, match={Intelligence}, replace={Intell.}]
			\step[fieldsource=\regexp{$MAPLOOP}, match={Hungary}, replace={Hung.}]
			\step[fieldsource=\regexp{$MAPLOOP}, match={Hungarian}, replace={Hung.}]
			\step[fieldsource=\regexp{$MAPLOOP}, match={Intelligent}, replace={Intell.}]
			\step[fieldsource=\regexp{$MAPLOOP}, match={Hydraulics}, replace={Hydraul.}]
			\step[fieldsource=\regexp{$MAPLOOP}, match={Interactions}, replace={Interact.}]
			\step[fieldsource=\regexp{$MAPLOOP}, match={Hydrology}, replace={Hydrol.}]
			\step[fieldsource=\regexp{$MAPLOOP}, match={Internationales}, replace={Int.}]
			\step[fieldsource=\regexp{$MAPLOOP}, match={International}, replace={Int.}]
			\step[fieldsource=\regexp{$MAPLOOP}, match={Illuminating}, replace={Illum.}]
			\step[fieldsource=\regexp{$MAPLOOP}, match={Isotopes}, replace={Isot.}]
			\step[fieldsource=\regexp{$MAPLOOP}, match={Imaging}, replace={Imag.}]
			\step[fieldsource=\regexp{$MAPLOOP}, match={Israel}, replace={Isr.}]
			\step[fieldsource=\regexp{$MAPLOOP}, match={Industrial}, replace={Ind.}]
			\step[fieldsource=\regexp{$MAPLOOP}, match={Japan}, replace={Jpn.}]
			\step[fieldsource=\regexp{$MAPLOOP}, match={Information}, replace={Inf.}]
			\step[fieldsource=\regexp{$MAPLOOP}, match={Journal}, replace={J.}]
			\step[fieldsource=\regexp{$MAPLOOP}, match={Informatics}, replace={Inform.}]
			\step[fieldsource=\regexp{$MAPLOOP}, match={Knowledge}, replace={Knowl.}]
			\step[fieldsource=\regexp{$MAPLOOP}, match={Laboratory}, replace={Lab.}]
			\step[fieldsource=\regexp{$MAPLOOP}, match={Laboratories}, replace={Lab.}]
			\step[fieldsource=\regexp{$MAPLOOP}, match={Mathematical}, replace={Math.}]
			\step[fieldsource=\regexp{$MAPLOOP}, match={Language}, replace={Lang.}]
			\step[fieldsource=\regexp{$MAPLOOP}, match={Mathematics}, replace={Math.}]
			\step[fieldsource=\regexp{$MAPLOOP}, match={Abstracts}, replace={Abstr.}]
			\step[fieldsource=\regexp{$MAPLOOP}, match={Analysis}, replace={Anal.}]
			\step[fieldsource=\regexp{$MAPLOOP}, match={Academy}, replace={Acad.}]
			\step[fieldsource=\regexp{$MAPLOOP}, match={Annals}, replace={Ann.}]
			\step[fieldsource=\regexp{$MAPLOOP}, match={Accelerator}, replace={Accel.}]
			\step[fieldsource=\regexp{$MAPLOOP}, match={Annual}, replace={Annu.}]
			\step[fieldsource=\regexp{$MAPLOOP}, match={Acoustics}, replace={Acoust.}]
			\step[fieldsource=\regexp{$MAPLOOP}, match={Apparatus}, replace={App.}]
			\step[fieldsource=\regexp{$MAPLOOP}, match={Active}, replace={Act.}]
			\step[fieldsource=\regexp{$MAPLOOP}, match={Applications}, replace={Appl.}]
			\step[fieldsource=\regexp{$MAPLOOP}, match={Administration}, replace={Admin.}]
			\step[fieldsource=\regexp{$MAPLOOP}, match={Applied}, replace={Appl.}]
			\step[fieldsource=\regexp{$MAPLOOP}, match={Administrative}, replace={Administ.}]
			\step[fieldsource=\regexp{$MAPLOOP}, match={Approximate}, replace={Approx.}]
			\step[fieldsource=\regexp{$MAPLOOP}, match={Advanced}, replace={Adv.}]
			\step[fieldsource=\regexp{$MAPLOOP}, match={Advances}, replace={Adv.}]
			\step[fieldsource=\regexp{$MAPLOOP}, match={Archives}, replace={Arch.}]
			\step[fieldsource=\regexp{$MAPLOOP}, match={Archive}, replace={Arch.}]
			\step[fieldsource=\regexp{$MAPLOOP}, match={Aeronautics}, replace={Aeronaut.}]
			\step[fieldsource=\regexp{$MAPLOOP}, match={Artificial}, replace={Artif.}]
			\step[fieldsource=\regexp{$MAPLOOP}, match={Aerospace}, replace={Aerosp.}]
			\step[fieldsource=\regexp{$MAPLOOP}, match={Assembly}, replace={Assem.}]
			\step[fieldsource=\regexp{$MAPLOOP}, match={Affective}, replace={Affect.}]
			\step[fieldsource=\regexp{$MAPLOOP}, match={Association}, replace={Assoc.}]
			\step[fieldsource=\regexp{$MAPLOOP}, match={Africa}, replace={Afr.}]
			\step[fieldsource=\regexp{$MAPLOOP}, match={African}, replace={Afr.}]
			\step[fieldsource=\regexp{$MAPLOOP}, match={Astronomy}, replace={Astron.}]
			\step[fieldsource=\regexp{$MAPLOOP}, match={Aircraft}, replace={Aircr.}]
			\step[fieldsource=\regexp{$MAPLOOP}, match={Astronautics}, replace={Astronaut.}]
			\step[fieldsource=\regexp{$MAPLOOP}, match={Algebraic}, replace={Algebr.}]
			\step[fieldsource=\regexp{$MAPLOOP}, match={Astrophysics}, replace={Astrophys.}]
			\step[fieldsource=\regexp{$MAPLOOP}, match={American}, replace={Amer.}]
			\step[fieldsource=\regexp{$MAPLOOP}, match={Atmosphere}, replace={Atmos.}]
			\step[fieldsource=\regexp{$MAPLOOP}, match={Atomic}, replace={At.}]
			\step[fieldsource=\regexp{$MAPLOOP}, match={Atoms}, replace={At.}]
			\step[fieldsource=\regexp{$MAPLOOP}, match={Broadcasting}, replace={Broadcast.}]
			\step[fieldsource=\regexp{$MAPLOOP}, match={Australasian}, replace={Australas.}]
			\step[fieldsource=\regexp{$MAPLOOP}, match={Bulletin}, replace={Bull.}]
			\step[fieldsource=\regexp{$MAPLOOP}, match={Australia}, replace={Aust.}]
			\step[fieldsource=\regexp{$MAPLOOP}, match={Bureau}, replace={Bur.}]
			\step[fieldsource=\regexp{$MAPLOOP}, match={{Automatic }}, replace={Autom.}]
			\step[fieldsource=\regexp{$MAPLOOP}, match={Business}, replace={Bus.}]
			\step[fieldsource=\regexp{$MAPLOOP}, match={Automation}, replace={Automat.}]
			\step[fieldsource=\regexp{$MAPLOOP}, match={Canadian}, replace={Can.}]
			\step[fieldsource=\regexp{$MAPLOOP}, match={Automotive}, replace={Automot.}]
			\step[fieldsource=\regexp{$MAPLOOP}, match={Automobiles}, replace={Automob.}]
			\step[fieldsource=\regexp{$MAPLOOP}, match={Automobile}, replace={Automob.}]
			\step[fieldsource=\regexp{$MAPLOOP}, match={Ceramic}, replace={Ceram.}]
			\step[fieldsource=\regexp{$MAPLOOP}, match={Autonomous}, replace={Auton.}]
			\step[fieldsource=\regexp{$MAPLOOP}, match={Chemical}, replace={Chem.}]
			\step[fieldsource=\regexp{$MAPLOOP}, match={Behavioral}, replace={Behav.}]
			\step[fieldsource=\regexp{$MAPLOOP}, match={Behavior}, replace={Behav.}]
			\step[fieldsource=\regexp{$MAPLOOP}, match={Chinese}, replace={Chin.}]
			\step[fieldsource=\regexp{$MAPLOOP}, match={Belgian}, replace={Belg.}]
			\step[fieldsource=\regexp{$MAPLOOP}, match={Climatology}, replace={Climatol.}]
			\step[fieldsource=\regexp{$MAPLOOP}, match={Biochemical}, replace={Biochem.}]
			\step[fieldsource=\regexp{$MAPLOOP}, match={Clinical}, replace={Clin.}]
			\step[fieldsource=\regexp{$MAPLOOP}, match={Bioinformatics}, replace={Bioinf.}]
			\step[fieldsource=\regexp{$MAPLOOP}, match={Cognitive}, replace={Cogn.}]
			\step[fieldsource=\regexp{$MAPLOOP}, match={Biology, Biological}, replace={Biol.}]
			\step[fieldsource=\regexp{$MAPLOOP}, match={Colloquium}, replace={Colloq.}]
			\step[fieldsource=\regexp{$MAPLOOP}, match={Kolloquium}, replace={Kolloq.}]
			\step[fieldsource=\regexp{$MAPLOOP}, match={Biomedical}, replace={Biomed.}]
			\step[fieldsource=\regexp{$MAPLOOP}, match={Communications}, replace={Commun.}]
			\step[fieldsource=\regexp{$MAPLOOP}, match={Communication}, replace={Commun.}]
			\step[fieldsource=\regexp{$MAPLOOP}, match={Biophysics}, replace={Biophys.}]
			\step[fieldsource=\regexp{$MAPLOOP}, match={Compatibility}, replace={Compat.}]
			\step[fieldsource=\regexp{$MAPLOOP}, match={British}, replace={Brit.}]
			\step[fieldsource=\regexp{$MAPLOOP}, match={Components}, replace={Compon.}]
			\step[fieldsource=\regexp{$MAPLOOP}, match={Component}, replace={Compon.}]
			\step[fieldsource=\regexp{$MAPLOOP}, match={Computational}, replace={Comput.}]
			\step[fieldsource=\regexp{$MAPLOOP}, match={Delivery}, replace={Del.}]
			\step[fieldsource=\regexp{$MAPLOOP}, match={Computers}, replace={Comput.}]
			\step[fieldsource=\regexp{$MAPLOOP}, match={Computer}, replace={Comput.}]
			\step[fieldsource=\regexp{$MAPLOOP}, match={Department}, replace={Dept.}]
			\step[fieldsource=\regexp{$MAPLOOP}, match={Computing}, replace={Comput.}]
			\step[fieldsource=\regexp{$MAPLOOP}, match={Design}, replace={Des.}]
			\step[fieldsource=\regexp{$MAPLOOP}, match={Condensed}, replace={Condens.}]
			\step[fieldsource=\regexp{$MAPLOOP}, match={Detector}, replace={Detect.}]
			\step[fieldsource=\regexp{$MAPLOOP}, match={Conferences}, replace={Conf.}]
			\step[fieldsource=\regexp{$MAPLOOP}, match={Conference}, replace={Conf.}]
			\step[fieldsource=\regexp{$MAPLOOP}, match={Development}, replace={Develop.}]
			\step[fieldsource=\regexp{$MAPLOOP}, match={Congress}, replace={Congr.}]
			\step[fieldsource=\regexp{$MAPLOOP}, match={Differential}, replace={Differ.}]
			\step[fieldsource=\regexp{$MAPLOOP}, match={Consumer}, replace={Consum.}]
			\step[fieldsource=\regexp{$MAPLOOP}, match={Digest}, replace={Dig.}] 
			\step[fieldsource=\regexp{$MAPLOOP}, match={{ of the }}, replace={{ }}]
			\step[fieldsource=\regexp{$MAPLOOP}, match={{ of }}, replace={{ }}]
			\step[fieldsource=\regexp{$MAPLOOP}, match={{Of }}, replace={{ }}]
			\step[fieldsource=\regexp{$MAPLOOP}, match={{ on }}, replace={{ }}]
			\step[fieldsource=\regexp{$MAPLOOP}, match={{ On }}, replace={{ }}]
			\step[fieldsource=\regexp{$MAPLOOP}, match={{On }}, replace={{ }}]
			\step[fieldsource=\regexp{$MAPLOOP}, match={{ in }}, replace={{ }}]
			\step[fieldsource=\regexp{$MAPLOOP}, match=\regexp{\\\x{26}}, replace={{and}}] %
			\step[fieldsource=\regexp{$MAPLOOP}, match={{, and }}, replace={{, }}]
			\step[fieldsource=\regexp{$MAPLOOP}, match={{, And }}, replace={{, }}]
			\step[fieldsource=\regexp{$MAPLOOP}, match={{ and }}, replace={{ }}]
			\step[fieldsource=\regexp{$MAPLOOP}, match={{ And }}, replace={{ }}]
			\step[fieldsource=\regexp{$MAPLOOP}, match={{ In }}, replace={{ }}]
			\step[fieldsource=\regexp{$MAPLOOP}, match={{In }}, replace={{ }}]
			\step[fieldsource=\regexp{$MAPLOOP}, match={{ the }}, replace={{ }}] 
			\step[fieldsource=\regexp{$MAPLOOP}, match={{ The }}, replace={{ }}] 
			\step[fieldsource=\regexp{$MAPLOOP}, match={{The }}, replace={}] 
			\step[fieldsource=\regexp{$MAPLOOP}, match={{ for }}, replace={{ }}] 
			\step[fieldsource=\regexp{$MAPLOOP}, match={First}, replace={1st}]
			\step[fieldsource=\regexp{$MAPLOOP}, match={Second}, replace={2nd}]
			\step[fieldsource=\regexp{$MAPLOOP}, match={Third}, replace={3rd}]
			\step[fieldsource=\regexp{$MAPLOOP}, match={Fourth}, replace={4th}]
			\step[fieldsource=\regexp{$MAPLOOP}, match={Fifth}, replace={5th}]
			\step[fieldsource=\regexp{$MAPLOOP}, match={Sixth}, replace={6th}]
			\step[fieldsource=\regexp{$MAPLOOP}, match={Seventh}, replace={7th}]
			\step[fieldsource=\regexp{$MAPLOOP}, match={Eighth}, replace={8th}]
			\step[fieldsource=\regexp{$MAPLOOP}, match={Ninth}, replace={9th}]
			\step[fieldsource=\regexp{$MAPLOOP}, match={Tenth}, replace={10th}]
			\step[fieldsource=\regexp{$MAPLOOP}, match={Eleventh}, replace={11th}]
			\step[fieldsource=\regexp{$MAPLOOP}, match={Twelfth}, replace={12th}]
			\step[fieldsource=\regexp{$MAPLOOP}, match={Thirteenth}, replace={13th}]
			\step[fieldsource=\regexp{$MAPLOOP}, match={Fourteenth}, replace={14th}]
			\step[fieldsource=\regexp{$MAPLOOP}, match={Fifteenth}, replace={15th}]
			\step[fieldsource=\regexp{$MAPLOOP}, match={Sixteenth}, replace={16th}]
			\step[fieldsource=\regexp{$MAPLOOP}, match={Seventeenth}, replace={17th}]
			\step[fieldsource=\regexp{$MAPLOOP}, match={Eighteenth}, replace={18th}]
			\step[fieldsource=\regexp{$MAPLOOP}, match={Nineteenth}, replace={19th}]
			\step[fieldsource=\regexp{$MAPLOOP}, match={Twentieth}, replace={20th}]
		}
	}
}

\usepackage{lipsum}
\usepackage{amsmath}
\usepackage{amssymb} 	%

\usepackage{amsthm} 	%

\newtheoremstyle{defn}%
	{3pt}%
	{3pt}%
	{\addtolength{\leftskip}{\parindent}\itshape}%
	{}%
	{\bfseries}%
	{:}%
	{.5em}%
	{\thmnote{#3} {\normalfont(\thmname{#1} \thmnumber{#2})}}%

\newtheoremstyle{term}%
	{3pt}%
	{3pt}%
	{\addtolength{\leftskip}{\parindent}\itshape}%
	{}%
	{\bfseries}%
	{:}%
	{.5em}%
	{\thmnote{#3}}%

\theoremstyle{defn}
\newtheorem{definition}{Definition}
\theoremstyle{term}
\newtheorem{term}{Term}
\usepackage{bm} 		%
\usepackage{booktabs} 	%
\usepackage{csquotes}
\usepackage{microtype}
\usepackage[all]{nowidow}
\usepackage{fmtcount}
\usepackage{stackengine}
\usepackage{flushend}
\usepackage[hidelinks]{hyperref}

\usepackage{hologo}
\usepackage{listings}
\usepackage{lstautogobble}
\newcommand{\myorcidlink}[1]{}

\ifx \isaccepted \undefined
\newcommand\copyrighttext{%
	\footnotesize \centering This work has been submitted to the IEEE for possible publication.\\ Copyright may be transferred without notice, after which this version may no longer be accessible.}
\else
\newcommand\copyrighttext{%
	\footnotesize \parbox[t]{.11\textwidth}{\copyright{} \the\year~IEEE.} \parbox[t]{.87\textwidth}{Personal use of this material is permitted. Permission from IEEE must be obtained for all other uses, in any current or future media, including reprinting/republishing this material for advertising or promotional purposes, creating new collective works, for resale or redistribution to servers or lists, or reuse of any copyrighted component of this work in other works.}}
\fi
\newcommand\copyrightnotice{%
	\ifx \compileforpublish \undefined
	\else
	\begin{tikzpicture}[remember picture,overlay]
	\node[anchor=south,yshift=50pt] at (current page.south) {\parbox{\dimexpr\textwidth-\fboxsep-\fboxrule\relax}{\copyrighttext}};
	\end{tikzpicture}%
	\fi
}

\renewcommand\vec{\mathbf}
\newcommand\ubar[1]{\stackunder[1.2pt]{$#1$}{\rule{.8ex}{.075ex}}}

\newcommand{\cf}{cf.}
\newcommand{\eg}{e.g.}
\newcommand{\ia}{i.a.}
\newcommand{\ie}{i.e.}

\newcommand{\Safestate}{\emph{Safe state}}
\newcommand{\safestate}{\emph{safe state}}
\newcommand{\Failop}{\emph{Fail-operational}}
\newcommand{\failop}{\emph{fail-operational}}
\newcommand{\Failsafe}{\emph{Fail-safe}}
\newcommand{\failsafe}{\emph{fail-safe}}

\newcommand{\failsilent}{\emph{fail-silent}}
\newcommand{\Faildegraded}{\emph{Fail-degraded}}
\newcommand{\faildegraded}{\emph{fail-degraded}}

\newcommand{\failreduced}{\emph{fail-reduced}}

\newcommand{\Failunsafe}{\emph{Fail-unsafe}}
\newcommand{\failunsafe}{\emph{fail-unsafe}}
\newcommand{\Op}{\emph{Operational}}
\newcommand{\op}{\emph{operational}}

\newcommand\circled[1]{\tikz[baseline=(char.base)]{\node[shape=circle,draw,inner sep=.5pt,fill=white] (char) {#1};}}

\definecolor{ourRed}{named}{tuRed}%
\colorlet{ourRed}{tuRed!30}
\definecolor{ourBlue}{named}{tuBlue}%
\colorlet{ourBlue}{tuBlue!30}

\begin{document}
\thispagestyle{empty}
\pagestyle{empty}
\twocolumn[
\begin{@twocolumnfalse}
	%
	%
	
	\Huge {IEEE copyright notice} \\ \\ 
	\large {\copyright\ 2021 IEEE. Personal use of this material is permitted. Permission from IEEE must be obtained for all other uses, in any current or future media, including reprinting/republishing this material for advertising or promotional purposes, creating new collective works, for resale or redistribution to servers or lists, or reuse of any copyrighted component of this work in other works.} \\ \\
	
	{\Large Published in IEEE Transactions on Intelligent Vehicles} \\ \\
	
	{\Large DOI: \href{https://doi.org/10.1109/TIV.2021.3129933}{10.1109/TIV.2021.3129933}} \\ \\
	
	Cite as:
	\vspace{0.1cm}
	
	\noindent\fbox{%
		\begin{minipage}{0.98\textwidth}%
			T.~Stolte, S.~Ackermann, R.~Graubohm, I.~Jatzkowski, B.~Klamann, H.~Winner, and M.~Maurer, ``A {Taxonomy} to {Unify} {Fault} {Tolerance} {Regimes} for {Automotive} {Systems}: {Defining} {Fail}-{Operational}, {Fail}-{Degraded}, and {Fail}-{Safe},'' IEEE Transactions on Intelligent Vehicles, vol.~7, no.~2, pp.~251–262, 2022. DOI: 10.1109/TIV.2021.3129933.
		\end{minipage}
	}
	\vspace{2cm}
	
\end{@twocolumnfalse}
]

\noindent%
\hologo{BibTeX}:

\noindent
	\begin{centering}
	\footnotesize
	\begin{lstlisting}[frame=single,linewidth=\textwidth]
		@article{stolte_2022,
			title = {A taxonomy to unify fault tolerance regimes for automotive systems: defining fail-operational, fail-degraded, and fail-safe},
			volume = {7},
			doi = {10.1109/TIV.2021.3129933},
			pages = {251 -- 262},
			number = {2},
			journaltitle = {{IEEE} Transactions on Intelligent Vehicles},
			author = {Stolte, Torben and Ackermann, Stefan and Graubohm, Robert and Jatzkowski, Inga and Klamann, Bj\"orn and Winner, Hermann and Maurer, Markus},
			date = {2022-07},
			langid = {english}
		}
	\end{lstlisting}
\end{centering}

\title{\LARGE \bf
	A Taxonomy to Unify Fault Tolerance Regimes for Automotive Systems: Defining Fail-Operational, Fail-Degraded, and Fail-Safe
}

\author{Torben Stolte\myorcidlink{0000-0003-3697-8046}, 
		Stefan Ackermann\myorcidlink{0000-0002-0984-0423}, 
		Robert Graubohm\myorcidlink{0000-0002-6682-4788}, 
		Inga Jatzkowski\myorcidlink{0000-0003-4127-3913},\\
		Björn Klamann\myorcidlink{0000-0002-2589-9812},
		Hermann Winner\myorcidlink{0000-0002-9824-3195}, and 
		Markus Maurer\myorcidlink{0000-0002-5357-9701}%
\thanks{%
		This research is accomplished within the project ``UNICAR\emph{agil}''~\cite{woopen_2018}~(FKZ 16EMO0285, FKZ 16EMO0286). 
		We acknowledge the financial support for the project by the German Federal Ministry of Education and Research~(BMBF).~%
		\emph{(Corresponding author: Torben Stolte)}}%
\thanks{Torben Stolte, Robert Graubohm, Inga Jatzkowski, and Markus Maurer are with the Institute of Control Engineering at TU Braunschweig, 38106 Braunschweig, Germany
        {\tt\small \{lastname\}@ifr.ing.tu-bs.de}}%
\thanks{Stefan Ackermann, Björn Klamann, and Hermann Winner are with the Institute of Automotive Engineering at TU Darmstadt, 64287 Darmstadt, Germany
        {\tt\small \{firstname(ö=oe).lastname\}@tu-darmstadt.de}}%
}

\markboth{Transactions on Intelligent Vehicles,~Vol.~AA, No.~B, MONTH~YEAR}%
{Stolte \MakeLowercase{\textit{et al.}}: A Taxonomy to Unify Fault Tolerance Regimes for Automotive Systems}

\maketitle

\begin{abstract}

This paper presents a taxonomy that allows defining the fault tolerance regimes \emph{fail-operational}, \emph{fail-degraded}, and \emph{fail-safe} in the context of automotive systems. 
Fault tolerance regimes such as these are widely used in recent publications related to automated driving, yet without definitions.
This largely holds true for automotive safety standards, too. 
We show that fault tolerance regimes defined in scientific publications related to the automotive domain are partially ambiguous as well as taxonomically unrelated. 
The presented taxonomy is based on terminology stemming from ISO\,26262 as well as from systems engineering.
It uses four criteria to distinguish fault tolerance regimes. %
In addition to \emph{fail-operational}, \emph{fail-degraded}, and \emph{fail-safe}, the core terminology consists of \emph{operational} and \emph{fail-unsafe}. 
These terms are supported by definitions of \emph{available performance}, \emph{nominal performance}, \emph{functionality}, and a concise definition of the \emph{safe state}.
For verification, we show by means of two examples from the automotive domain that the taxonomy can be applied to hierarchical systems of different complexity.

\end{abstract}%

\IEEEpeerreviewmaketitle

\section{INTRODUCTION}

\newcommand{\citeFailOp}{\cite{benz_2004,carre_2020,chen_2008,gleirscher_2019,isermann_2000,isermann_2002,iso_2020,li_2019,martinus_2004,mauritz_2019,messnarz_2019,reif_2014,schauffele_2016,schmid_2019,schnellbach_2016,schnellbach_2016a,thorn_2018,wanner_2012a,wood_2019}}
\newcommand{\citeFailSafe}{\cite{benz_2004,carre_2020,chen_2008,isermann_2000,isermann_2002,iso_2020,li_2019,mauritz_2019,messnarz_2019,schauffele_2016,schmid_2019,schnellbach_2016,schnellbach_2016a,stetter_2020,thorn_2018,wanner_2012a,wood_2019}}
\newcommand{\citeFailSilent}{\cite{benz_2004,carre_2020,chen_2008,gleirscher_2019,isermann_2000,isermann_2002,li_2019,martinus_2004,reif_2014,schmid_2019,schnellbach_2016a,wanner_2012a}}
\newcommand{\citeFailSoft}{\cite{}}
\newcommand{\citeFailDegraded}{\cite{chen_2008,iso_2020,wood_2019}}
\newcommand{\citeFailReduced}{\cite{reif_2014,schauffele_2016}}
\newcommand{\citeFailOver}{\cite{}}
\newcommand{\citeFailArbitrary}{\cite{}}
\newcommand{\citeExamplesWDefinition}{\cite{benz_2004,carre_2020,chen_2008,gleirscher_2019,isermann_2000,isermann_2002,iso_2020,li_2019,martinus_2004,mauritz_2019,messnarz_2019,reif_2014,schauffele_2016,schmid_2019,schnellbach_2016,schnellbach_2016a,stetter_2020,thorn_2018,wanner_2012a,wood_2019}}
\newcommand{\citeExamplesWODefinition}{\cite{adler_2019,bartels_2015,becker_2015,becker_2017,bertino_2019,beyerer_2019,bijlsma_2017,fruehling_2019,goth_2020,helmle_2014,klomp_2019,magdici_2016,matute-peaspan_2020,mostl_2016,niedballa_2020,ramanathanvenkita_2020,sari_2020,sinha_2011,stolte_2016,weiss_2016,weiss_2016a,witte_2017}}
\newcommand{\printHeader}{
\node [header] at (M-1-1.north) {Author(s)};
\node [header] at (M-1-2.north) {Year};
\node [rotate=90, anchor=west] at (M-1-3.north)(node5){Source};
\node [rotate=60, anchor=west] at (M-1-4.north)(node6){fail-operational};
\node [rotate=60, anchor=west] at (M-1-5.north)(node7){fail-safe};
\node [rotate=60, anchor=west] at (M-1-6.north)(node8){fail-silent};
\node [rotate=60, anchor=west] at (M-1-7.north)(node9){fail-degraded};
\node [rotate=60, anchor=west] at (M-1-8.north)(node10){fail-reduced};
\node [rotate=60, anchor=west] at (M-1-9.north)(node11){fail-unsafe};
\node [rotate=60, anchor=west] at (M-1-10.north) {other};
}
\def\lastRowWODefinition{22}
\def\firstRowWDefinition{23}

\IEEEPARstart{I}{n} complex safety-critical systems, fault tolerance is a crucial property to ensure operation at an acceptable risk level.
The complexity of mechatronic systems such as vehicle systems allows a distinction to be made between different forms of fault tolerance.
These are often classified by developers and researchers in regimes such as \failop, \faildegraded, or \failsafe.
However, technical literature and relevant standards show a lack of uniform definitions for these regimes.
This leads to the frequent occurrence of different interpretations and designations of existing concepts; many publications in the context of automated vehicles even use the designations of fault tolerance regimes without stating or referring to any definition~\citeExamplesWODefinition, as presented in \autoref{tab:overview}.

Yet, a uniform understanding of fault tolerance regimes is essential both for the scientific discussion of fault tolerance in vehicle systems and for the communication between interdisciplinary developers in distributed development projects and other important stakeholders.
In particular, the concrete specification of fault tolerance, and therefore a consistent use of terminology becomes even more important when it comes to the development of and safety argumentation for SAE Level~4+~\cite{sae_2021} automated vehicles. 
Vehicle automation impedes a safety argumentation because drivers cannot be included as a fallback layer in the safety concept of the automated driving functionality. 
Hence, the use of the addressed terms in the literature with a focus on the automotive domain is presented in \autoref{sec:stateoftheart}.
In \autoref{sec:taxonomy}, we propose a taxonomy for fault tolerance regimes that can be applied coherently to automotive systems. 
The applicability at different architectural levels is demonstrated by two automotive examples in \autoref{sec:application}.
\autoref{sec:application} also contains a verification of a set of requirements.

\section{STATE OF THE ART}
\label{sec:stateoftheart}

In order to illustrate the understanding of fault tolerance regimes in literature, 
we summarize publications that introduce definitions in an automotive context in \autoref{subsec:deviatinguse} and additionally present selected publications from other domains in \autoref{subsec:otherdomains}.
In the following subsections, terms in direct and indirect quotations are used according to the understanding of the cited authors. Terms in our own statements are in accordance with the definitions that we present in \autoref{sec:taxonomy}.

\newlength{\headerHeight}
\setlength{\headerHeight}{\heightof{\footnotesize ()}}
\newlength{\headerDepth}
\setlength{\headerDepth}{\depthof{\footnotesize ()}}

\newlength{\cellwidth}
\setlength{\cellwidth}{\totalheightof{\footnotesize fp}}
\newlength{\colseparator}
\setlength{\colseparator}{0.125em}
\newlength{\yearwidth}
\setlength{\yearwidth}{\widthof{\footnotesize (2020)}}
\newlength{\sourcewidth}
\setlength{\sourcewidth}{\widthof{\footnotesize [66]}}

\newlength{\authorwidth}
\setlength{\authorwidth}{\linewidth}
\addtolength{\authorwidth}{-12.0\cellwidth}
\addtolength{\authorwidth}{-\yearwidth}
\addtolength{\authorwidth}{-\sourcewidth}
\addtolength{\authorwidth}{-17.0\colseparator}
\addtolength{\authorwidth}{-.5cm}

\newlength{\firstrowheight}
\setlength{\firstrowheight}{\widthof{\footnotesize fail-operational}}
\newlength{\nodeheight}
\setlength{\nodeheight}{\totalheightof{\footnotesize ()}}

\newlength{\rowseparator}
\setlength{\rowseparator}{.05cm}

\makeatletter
\tikzset{
	store number of columns in/.style={execute at end matrix={
			\xdef#1{\the\pgf@matrix@numberofcolumns}}},
	store number of rows in/.style={execute at end matrix={
			\xdef#1{\the\pgfmatrixcurrentrow}}}}
\makeatother

\newcommand{\kreisvoll}{\begin{tikzpicture}\draw[fill=black](0,0)circle(.5mm);\end{tikzpicture}}

\begin{table}[t]
	\caption{A non-exhaustive overview of publications using fault tolerance regimes in an automotive context.}
	\label{tab:overview}
	\centering
	\begin{tikzpicture}
		\footnotesize
		\tikzstyle{column 1}=[text width=\authorwidth, align=right,draw,inner sep=0]
		\tikzstyle{column 2}=[minimum width=\yearwidth]
		\tikzstyle{column 3}=[minimum width=\sourcewidth]
		\tikzstyle{column 4}=[minimum width=\cellwidth]
		\tikzstyle{column 5}=[minimum width=\cellwidth]
		\tikzstyle{column 6}=[minimum width=\cellwidth]
		\tikzstyle{column 7}=[minimum width=\cellwidth]
		\tikzstyle{column 8}=[minimum width=\cellwidth]
		\tikzstyle{column 9}=[minimum width=\cellwidth]
		\tikzstyle{column 10}=[minimum width=\cellwidth]
		\tikzstyle{column 11}=[minimum width=\cellwidth]
		\tikzstyle{column 12}=[minimum width=\cellwidth]
		\tikzstyle{column 13}=[minimum width=\cellwidth]
		\tikzstyle{column 14}=[minimum width=\cellwidth]
		\tikzstyle{column 15}=[minimum width=\cellwidth]
		\tikzstyle{column 16}=[minimum width=\cellwidth]
		\tikzstyle{header} = [text height = \headerHeight, text depth = \headerDepth, inner sep =0, anchor =south]
		\matrix (M) [	matrix of nodes,
						nodes in empty cells,
						column sep={\colseparator},
						row sep = \rowseparator,
						nodes={	align=center,
								inner sep = 0, 
								outer sep = 0,
								anchor=center,
								minimum height = \nodeheight
							},
						store number of columns in=\colcount,
						store number of rows in=\rowcount
					]
		{%
			\citeauthor{adler_2019}&(\citeyear{adler_2019})&\cite{adler_2019}&\kreisvoll&\kreisvoll&&&&&\\
\citeauthor{bartels_2015}&(\citeyear{bartels_2015})&\cite{bartels_2015}&\kreisvoll&\kreisvoll&&&&&\\
\citeauthor{becker_2015}&(\citeyear{becker_2015})&\cite{becker_2015}&\kreisvoll&&&\kreisvoll&&&\\
\citeauthor{becker_2017}&(\citeyear{becker_2017})&\cite{becker_2017}&\kreisvoll&&&\kreisvoll&&&\\
\citeauthor{bertino_2019}&(\citeyear{bertino_2019})&\cite{bertino_2019}&\kreisvoll&&&&&&\\
\citeauthor{beyerer_2019}&(\citeyear{beyerer_2019})&\cite{beyerer_2019}&&\kreisvoll&&&&&\kreisvoll\\
\citeauthor{bijlsma_2017}&(\citeyear{bijlsma_2017})&\cite{bijlsma_2017}&\kreisvoll&\kreisvoll&&&&&\\
\citeauthor{fruehling_2019}&(\citeyear{fruehling_2019})&\cite{fruehling_2019}&\kreisvoll&\kreisvoll&&&&&\\
\citeauthor{goth_2020}&(\citeyear{goth_2020})&\cite{goth_2020}&\kreisvoll&\kreisvoll&&&&\kreisvoll&\\
\citeauthor{helmle_2014}&(\citeyear{helmle_2014})&\cite{helmle_2014}&\kreisvoll&&&\kreisvoll&&&\\
\citeauthor{klomp_2019}&(\citeyear{klomp_2019})&\cite{klomp_2019}&\kreisvoll&\kreisvoll&\kreisvoll&&&&\\
\citeauthor{magdici_2016}&(\citeyear{magdici_2016})&\cite{magdici_2016}&&\kreisvoll&&&&&\\
\citeauthor{matute-peaspan_2020}&(\citeyear{matute-peaspan_2020})&\cite{matute-peaspan_2020}&\kreisvoll&&&&&&\\
\citeauthor{mostl_2016}&(\citeyear{mostl_2016})&\cite{mostl_2016}&\kreisvoll&\kreisvoll&&&&&\\
\citeauthor{niedballa_2020}&(\citeyear{niedballa_2020})&\cite{niedballa_2020}&\kreisvoll&\kreisvoll&&&&&\kreisvoll\\
\citeauthor{ramanathanvenkita_2020}&(\citeyear{ramanathanvenkita_2020})&\cite{ramanathanvenkita_2020}&\kreisvoll&&&&&&\\
\citeauthor{sari_2020}&(\citeyear{sari_2020})&\cite{sari_2020}&\kreisvoll&\kreisvoll&&&&&\\
\citeauthor{sinha_2011}&(\citeyear{sinha_2011})&\cite{sinha_2011}&\kreisvoll&&\kreisvoll&&&&\\
\citeauthor{stolte_2016}&(\citeyear{stolte_2016})&\cite{stolte_2016}&\kreisvoll&\kreisvoll&&&&&\\
\citeauthor{weiss_2016}&(\citeyear{weiss_2016})&\cite{weiss_2016}&\kreisvoll&\kreisvoll&\kreisvoll&&&&\\
\citeauthor{weiss_2016a}&(\citeyear{weiss_2016a})&\cite{weiss_2016a}&\kreisvoll&\kreisvoll&\kreisvoll&&&&\\
\citeauthor{witte_2017}&(\citeyear{witte_2017})&\cite{witte_2017}&\kreisvoll&\kreisvoll&&&&&\\
\citeauthor{benz_2004}&(\citeyear{benz_2004})&\cite{benz_2004}&\kreisvoll&\kreisvoll&\kreisvoll&&&&\\
\citeauthor{carre_2020}&(\citeyear{carre_2020})&\cite{carre_2020}&\kreisvoll&\kreisvoll&\kreisvoll&&&&\\
\citeauthor{chen_2008}&(\citeyear{chen_2008})&\cite{chen_2008}&\kreisvoll&\kreisvoll&\kreisvoll&\kreisvoll&&&\\
\citeauthor{gleirscher_2019}&(\citeyear{gleirscher_2019})&\cite{gleirscher_2019}&\kreisvoll&&\kreisvoll&&&&\\
\citeauthor{isermann_2000}&(\citeyear{isermann_2000})&\cite{isermann_2000}&\kreisvoll&\kreisvoll&\kreisvoll&&&&\\
\citeauthor{isermann_2002}&(\citeyear{isermann_2002})&\cite{isermann_2002}&\kreisvoll&\kreisvoll&\kreisvoll&&&&\\
ISO/TR\,4804&(\citeyear{iso_2020})&\cite{iso_2020}&\kreisvoll&\kreisvoll&&\kreisvoll&&&\\
\citeauthor{li_2019}&(\citeyear{li_2019})&\cite{li_2019}&\kreisvoll&\kreisvoll&\kreisvoll&&&&\\
\citeauthor{martinus_2004}&(\citeyear{martinus_2004})&\cite{martinus_2004}&\kreisvoll&&\kreisvoll&&&&\\
\citeauthor{mauritz_2019}&(\citeyear{mauritz_2019})&\cite{mauritz_2019}&\kreisvoll&\kreisvoll&&&&&\\
\citeauthor{messnarz_2019}&(\citeyear{messnarz_2019})&\cite{messnarz_2019}&\kreisvoll&\kreisvoll&&&&&\\
\citeauthor{reif_2014}&(\citeyear{reif_2014})&\cite{reif_2014}&\kreisvoll&&\kreisvoll&&\kreisvoll&&\\
\citeauthor{schauffele_2016}&(\citeyear{schauffele_2016})&\cite{schauffele_2016}&\kreisvoll&\kreisvoll&&&\kreisvoll&&\\
\citeauthor{schmid_2019}&(\citeyear{schmid_2019})&\cite{schmid_2019}&\kreisvoll&\kreisvoll&\kreisvoll&&&&\\
\citeauthor{schnellbach_2016}&(\citeyear{schnellbach_2016})&\cite{schnellbach_2016}&\kreisvoll&\kreisvoll&&&&&\\
\citeauthor{schnellbach_2016a}&(\citeyear{schnellbach_2016a})&\cite{schnellbach_2016a}&\kreisvoll&\kreisvoll&\kreisvoll&&&&\\
\citeauthor{stetter_2020}&(\citeyear{stetter_2020})&\cite{stetter_2020}&&\kreisvoll&&&&&\kreisvoll\\
\citeauthor{thorn_2018}&(\citeyear{thorn_2018})&\cite{thorn_2018}&\kreisvoll&\kreisvoll&&&&&\\
\citeauthor{wanner_2012a}&(\citeyear{wanner_2012a})&\cite{wanner_2012a}&\kreisvoll&\kreisvoll&\kreisvoll&&&&\\
\citeauthor{wood_2019}&(\citeyear{wood_2019})&\cite{wood_2019}&\kreisvoll&\kreisvoll&&\kreisvoll&&&\\%
		};
		\foreach \i in {4,...,\colcount}
		{
			\draw (M-1-\i.north)--(M-\rowcount-\i.center);
		}
		\foreach \i in {1,...,\rowcount}
		{
			\draw ($(M-\i-3.east)+(.05cm,0cm)$)--(M-\i-3-|M-\i-\colcount.center);
			\draw ($(M-\i-3.north east)+(.05cm,0cm)$)--($(M-\i-3.south east)+(.05cm,0cm)$);
		}
		\printHeader
		\draw [decorate,decoration={brace,amplitude=5pt,raise=0pt},yshift=0pt] (M-\rowcount-1.south west) -- node [midway, above, rotate=90,yshift=3pt, align=center]{Automotive publications with defintions\\of fault tolerance regimes} (M-\firstRowWDefinition-1.north west);
		\draw [decorate,decoration={brace,amplitude=5pt,raise=0pt},yshift=0pt] (M-\lastRowWODefinition-1.south west) -- node [midway, above, rotate=90,yshift=3pt, align=center]{Publications in automated vehicle context\\using fault tolerance regimes without definitions} (M-1-1.north west);
		\draw (M-\rowcount-\colcount.west)-|(M-\rowcount-\colcount.north);
	\end{tikzpicture}
\end{table}

\subsection{Fault Tolerance Regimes in the Automotive Domain}
\label{subsec:deviatinguse}

In the automotive domain, standards such as the recent versions of ISO\,26262~\cite{iso_2018} or ISO/DIS\,21448~\cite{iso_2021} introduce extensive safety-related terminology.
However, the terminology in these two standards does not include definitions of fault tolerance regimes. 
The same applies to SAE~J3016~\cite{sae_2021}.
An understanding of \failsafe\ at the vehicle level is described by the UNECE in~\cite{ECE_2020} for automated driving applications, yet without distinguishing other fault tolerance regimes.

Aside from standards, several publications related to the automotive domain give definitions for different fault tolerance regimes~\citeExamplesWDefinition.
An overview of the covered literature is presented in \autoref{tab:overview}. 
In general, the terms found in literature can be divided into three groups based on the desired system behavior in the presence of a fault.
The first group consists of terms that indicate that a system can provide its specified functionality even in the presence of a fault.
Terms that target systems that provide impaired functionality after a fault make up the second group.
Terms in the third group require systems to fail into a defined state.

\subsubsection{Upholding functionality}
\label{subsubsec:failop}

With one exception, every automotive publication with definitions of fault tolerance regimes introduced in \autoref{tab:overview} includes a regime describing the upholding of functionality~\citeFailOp.
These publications define terms to address the continued provision of a system's functionality in the presence of a fault without performance degradation and consistently use the term \failop.
Still, when comparing the definitions, the understanding of the term varies slightly between the publications. 

The publications defining \failop\ can be divided into three main categories. 
The first main category includes all publications that expect a \failop\ system to strive towards achieving a defined state in the event of a fault and will abort normal operation. 
The authors of the cited publications refer to the defined state almost exclusively as the safe state. 
This understanding of the safe state does not match the definition we present in this paper.
This first main category can be divided into two subcategories. 
In the first subcategory~\cite{benz_2004, gleirscher_2014, luo_2019a, thorn_2018,messnarz_2019,matute-peaspan_2020}, a core functionality of the \failop\ system is maintained to achieve a defined state.
The sole publication of the second subcategory~\cite{gleirscher_2019} requires the system to maintain full functionality until the defined state is reached. 
It is not apparent to which subcategory each of the remaining publications~\cite{isermann_2000, isermann_2002, schauffele_2016, iso_2020} can be assigned as the provided definitions are not specific enough to make this distinction.

The publications of the second main category~\cite{chen_2008, martinus_2004, reif_2014, wanner_2012a} describe a \failop\ system as a system that maintains its full functionality despite a fault. 
The authors of the aforementioned publications do not expect a \failop\ system to achieve a defined state in case of a fault, but expect it to continue its normal operation.
The sole publication that is part of the third main category~\cite{wood_2019} deviates from the concept of maintaining functionality or achieving a defined state and rather describes a \failop\ system as a system that shall not lead to a safety-related situation in the event of a fault.

\subsubsection{Upholding functionality with reduced performance}

Researchers either use the term \faildegraded~\citeFailDegraded\ or \failreduced~\citeFailReduced\ to describe a reduced system performance while maintaining the system's functionality in the presence of a fault.
These terms are seldom defined in comparison to those describing either upholding a functionality or switching to a defined state. 

According to \citet{wood_2019} \faildegraded\ means \textquote[][]{[...] that the system is still able to operate safely when degraded.}
In contrast,~\citet{chen_2008} gives a more concise definition. 
He understands \faildegraded\ as the property of a system \textquote{which has the ability to continue with intended degraded operation at its output interfaces, despite the presence of hardware or software fault, [...].}
Thus,~\citeauthor{chen_2008} emphasizes two requirements for a system to be \faildegraded: (1) the system must continue its operation to be \faildegraded\ and (2) the continuation must follow a defined manner.
Showing a similar understanding, ISO/TR\,4804 defines \faildegraded\ as a property at the vehicle level when automated driving systems \textquote{operate with reduced functionality in the presence of a fault}~\cite{iso_2020}.

For~\citet{reif_2014}, a \failreduced\ system transitions into a state with a reduced functional capability in the presence of a fault. 
Similarly,~\citet{schauffele_2016} define \failreduced\ as a \textquote[][]{continued~-- albeit restricted --~system serviceability} in case of a fault.
Still,~\citet{reif_2014} as well as~\citet{schauffele_2016} do not specify further what is meant by ``reduced functional capability'' or ``restricted system serviceability,'' respectively.

A comparison of the definitions of \faildegraded\ and \failreduced\ reveals a very similar understanding among the authors. 
Either definition represents an understanding that a system is able to continue its intended functionality, yet with degraded performance.
Still, it becomes obvious that the understanding overlaps with the understanding that researchers have of \failop.

\subsubsection{Switching to a defined state}
Within the third group of terms, several authors use the terms \failsafe~\citeFailSafe\ and \failsilent~\citeFailSilent, either exclusively or in combination. 

The \failsafe\ property of a system is commonly described as the transition into a defined state (usually referred to as ``safe state'') in the event of failures~\cite{benz_2004,chen_2008,thorn_2018,isermann_2002,luo_2019a,reif_2014,schauffele_2016,wanner_2012a,wood_2019,messnarz_2019}.
While most definitions describe the safe state as a specific condition of the analyzed (sub-)system~\cite{benz_2004,chen_2008,isermann_2002,luo_2019a,reif_2014,schauffele_2016,wanner_2012a,messnarz_2019}, \citet[p.~90]{thorn_2018} argue that the safe state is a ``condition where the vehicle and occupants are safe.'' 
This corresponds to~\citet{wood_2019} who describe a fail-safe system to continue operating ``in a safe state in the event of a failure''~\cite[p.~135]{wood_2019}.
In~\cite{wood_2019}, it is not completely clear how the authors distinguish between \faildegraded\ and \failsafe\ as the understanding of the term \emph{system} is not further specified, \eg, whether system refers to the overall vehicle system or to different system levels within the vehicle.
\citet[p.~228]{luo_2019a} append that a system that reverts to a safe state generally no longer provides its required functionality.
\citet[p.~69]{isermann_2002} and~\citet[p.~599]{wanner_2012a} include in their definition that a fail-safe system can also be brought to a safe state externally or passively in the event of failures.
Finally,~\citet[p.~109]{schauffele_2016} and~\citet[p.~275]{reif_2014} stress the fact that, after transitioning to a safe state, a fail-safe system also has to maintain this safe state and exit it only after additional measures were taken (\eg, external reset).
The definition of the fail-safe property of an automated driving system in the technical report ISO/TR\,4804~\cite{iso_2020} specifies the need to achieve a \emph{minimal risk condition} in addition to a safe state in the event of a failure. 
This extension is largely consistent with the description of a ``Failsafe Response'' by the UNECE~\cite{ECE_2020}.

The \failsilent\ property of a system is commonly described as the guarantee that no system output is provided in the event of failures~\cite{benz_2004,chen_2008,gleirscher_2014,gleirscher_2019,isermann_2002,martinus_2004,reif_2014,wanner_2012a}. 
Therefore, the systems described as fail-silent usually represent \mbox{(sub-)}\-com\-po\-nents of a larger complex system (\eg, a vehicle).
Many authors name a complete shutdown or disabling of the communication of the system under consideration as a potential measure to achieve fail-silent behavior. 
Hence, the subsystems' functionality is no longer available in a larger system context, unless redundancies exist (\cf~\ref{subsubsec:failop}). 
\citet[p.~276]{reif_2014} appends that fail-silent systems should not only stop providing output signals, but should also stop reacting on any input signals after a failure occurred. 
\citet[p.~9]{chen_2008},~\citet[pp.~5]{gleirscher_2019}, and~\citet[p.~33]{martinus_2004} argue that fail-silent behavior is a possible manifestation of the fail-safe property of a system~\cite{martinus_2004}. %
\citet{carre_2020} provides conflicting explanations of the fail-silent property of a system. 
On the one hand, he explains a fail-silent component to be designed to ``continue operating properly in the event of the failure into a graceful degraded mode''~\cite[p.~56]{carre_2020}. 
On the other hand, he characterizes fail-silent behavior of a system by ``discontinued operation''~\cite[p.~66]{carre_2020}.
Deviating from other related work,~\citet[p.~104]{mauritz_2019} describes the fail-safe property of a component to be characterized by prevention ``from further interaction with the remaining system,'' which corresponds to the common definition of fail-silent behavior. 
The same can be observed in the work of \citet[p.~52]{stetter_2020}, who bases his explanation of the fail-safe strategy in system design (``enabling a controlled shut-down'' in the event of critical faults) on \citet{blanke_2000}.

\subsection{Definitions from Non-Automotive Domains}
\label{subsec:otherdomains}
Similar to the automotive-related publications discussed in \autoref{subsec:deviatinguse}, an inconsistent understanding of fault tolerance regimes can also be observed in other domains. 

Outlining concepts in the domain of dependable and secure computing,~\citet{avizienis_2004} use terms like \failsilent\ and \failsafe\ to describe a system's failure behavior. 
In the context of computing, the authors investigate \emph{service failure modes} that are described as the manifestation of a deviation from correct service~\cite[3.3.1]{avizienis_2004}. 
If, by design, a system displays only a specific failure mode,~\citet{avizienis_2004} call it a \emph{fail-controlled system}. 
The authors differentiate between various failure manifestations: 
They call a system with halted service failures a \emph{fail-halt system}, a system with stuck service and silent failures a \emph{fail-passive and fail-silent system}, and a system with minor failures (\ie, insignificant consequences) a \emph{fail-safe system}. 

Also with a background in dependable computing, \citeauthor{knight_2012} describes the \emph{fail-safe} property in~\cite{knight_2012}. 
The author points out that even a \emph{fail-safe} system that shuts down on error detection still exhibits a benign type of \emph{continued service} by falling and remaining silent~\cite[p.~131]{knight_2012}.

\citet{kopetz_2011} describes the use of \failsafe\ and \failop\ in real-time systems design. 
The author classifies real-time systems as \failsafe\ if they are able to detect failures and subsequently identify and quickly reach a safe state~\cite[p.~15]{kopetz_2011}. 
Consistent with some automotive publications discussed in the previous section,~\citeauthor{kopetz_2011} points out that a safe state is a condition of a controlled object and not of the designed computer system itself. 
In contrast to some publications from the automotive domain, the author classifies applications as \failop\ if they ``remain operational and provide a minimal level of service even in the case of a failure''~\cite[p.~15]{kopetz_2011}.

In the domain of nuclear safety, \failsafe\ is described by the \citet[11]{iaea_1991} as the behavior after a component or system failure, leading directly to a safe condition. 
It is further said that a component or system is only \failsafe\ for a stated kind of failure and situation. 
A contradicting understanding is used by \citet{moller_2008} while still referring to~\cite{iaea_1991}.
They use the term \emph{safe fail} instead of \failsafe\ to describe a system behaving safely after a component or system failure.
The term \failsafe\ is instead used to describe a system that is ``designed not to fail.''
\citet{moller_2008} also introduce the terms \emph{fail-silence} and \failop. 
They define \emph{fail-silence} consistently to the automotive domain as a mechanism that shuts down the system in case of a component failure.
They understand \emph{fail-silence} as a sub-category of a \emph{safe fail} mechanism.
In contrast, \failop\ is understood as a system that continues to work despite a fault occurrence by \citeauthor{moller_2008}. 
They also state that a distinction is sometimes made in related literature between the system remaining partially operational, which is then called \emph{fail-active}, and the system remaining fully operational.

\citet{nasa_2019} uses the term \failsafe\ in a slightly different way by defining it not only as an ability that safely terminates an operation, but potentially control the operation further after failure occurrence.

Finally, in the domain of fault-tolerant control systems,~\citet{blanke_2000,blanke_2016} present two definitions each for \failop\ and \failsafe\ as system-wide properties.
In~\cite{blanke_2000}, a \failop\ system \textquote[][]{is able to operate with no change in objectives or performance despite of any single failure,} while a \failsafe\ system is understood as a system that \textquote[][]{fails to a state that is considered safe in the particular context.}
The definition of \failop\ in~\cite[663]{blanke_2016} (\textquote[][]{The ability to sustain any single failure.}) is similar to~\cite{blanke_2000}.
In contrast, the definition of \failsafe\ in~\cite[662]{blanke_2016} appears inconclusive.

\subsection{Summary}
\label{subsec:summary}

Overall, the literature that uses and defines fault tolerance regimes shows a similar understanding with respect to the different regimes. 
Still, the understanding is not free of contradictions since slightly varying terms as well as a semantic overlap between the understanding of the terms can be observed in the different publications.
Moreover, the literature known to us does not provide taxonomic support for a clear distinction between fault tolerance regimes. 
Another aspect that is only partially addressed in the available literature is consistency in the use of related terms.
An exception is the technical report ISO/TR\,4804~\cite{iso_2020}, which embeds its definitions into the terminology defined in ISO\,26262~\cite{iso_2018}, ISO/DIS\,21448~\cite{iso_2021}, and SAE~J3016~\cite{sae_2021}.
However, ISO/TR\,4804~\cite{iso_2020} does not provide a taxonomic demarcation within the terminology and limits its definitions to the case of an automated driving system as the design item.
In contrast, fault tolerance regimes defined in other publications are applicable to arbitrary systems and system levels.

\section{TAXONOMY}

\label{sec:taxonomy}

\begin{figure*}
	\centering
	\input{terms.tex}
	\includegraphics{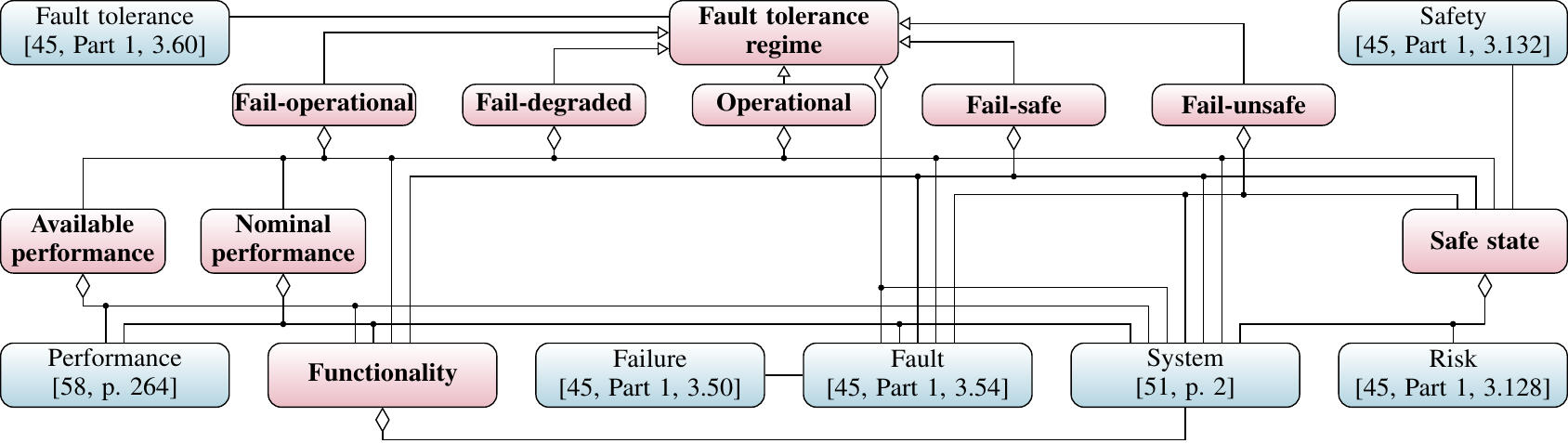}
	\caption{Terms defined in this paper (in bold) and related terms, where 
		\protect\tikz[baseline=-0.5ex]\protect\draw[aggregation](.35,0)--(0,0); indicates required terms, 	
		\protect\tikz[baseline=-0.5ex]\protect\draw[instance](0,0)--(.35,0); instances, and 
		\protect\tikz[baseline=-0.5ex]\protect\draw[relation](0,0)--(.35,0); semantically related terms.}
	\label{fig:terms}
\end{figure*}

The review of publications in \autoref{sec:stateoftheart} reveals a divergent understanding of fault tolerance regimes. 
Moreover, there are no established normative definitions for fault tolerance regimes available. 
However, for an interdisciplinary safety argumentation, a harmonized understanding of fault tolerances of complex systems is essential. 
We therefore present in this section a taxonomy in order to support a unified understanding of fault tolerance regime.
Subsection \ref{subsec:requirements} contains the requirements we used as basis for deriving the taxonomy.
In \autoref{subsec:relatedterms}, we describe the related terms on which the definitions of fault tolerance regimes rely before outlining the taxonomy in \autoref{subsec:definitions}.
\autoref{fig:terms} illustrates the newly defined terms, the related terms, as well as their interconnection.

\subsection{Requirements}
\label{subsec:requirements}

The new definitions shall support researchers and developers, who decide which safety functions are needed for specific systems or components. 
This goal leads to a set of requirements to reach a common understanding of the different fault tolerance regimes. 

To reach this common understanding, the new definitions need to be unambiguously understandable for all stakeholders of a domain. 
Therefore, the fault tolerance regimes have to be clearly separable from each other and have to resolve existing contradictions in current definitions as shown in \autoref{sec:stateoftheart}:

\begin{RQ}
	\item The fault tolerance regimes must be clearly distinguishable from each other by means of the taxonomy.%
	\label{req:distinguishable}
\end{RQ}

Furthermore, the new definitions need to be widely accepted in the community so that a majority of the stakeholders is aware of the definitions. 
Thus, the definitions need to be compatible to generally accepted definitions in existing literature as presented in \autoref{sec:stateoftheart}, which leads to the second requirement for our definitions:
\begin{RQ}[resume]
	\item The terms used for the fault tolerance regimes as well as the corresponding definitions should reflect the most common usage in the literature.%
	\label{req:compatibility}
\end{RQ}
\noindent Deviations from this may occur for the case that these definitions contradict a new logical argumentation of a novel definition.

Our focus in this paper is the application of the fault tolerance regime taxonomy to the automotive domain (in particular in the context of the automotive safety standards for electrical and electronic (E/E) systems ISO\,26262~\cite{iso_2018} and ISO/DIS\,21448~\cite{iso_2021}), bringing forth requirement~\ref{req:automotivedomain}:
\begin{RQ}[resume]
	\item The fault tolerance regime definitions must be applicable to systems of the automotive domain.%
	\label{req:automotivedomain}
\end{RQ}

In \autoref{sec:stateoftheart}, we point out that fault tolerance regimes are used at different system levels.
Therefore, the definitions should not only be applicable to the vehicle level, but should also cover the fault tolerance behavior of, \eg, subsystems. 
This means that different behavior at different levels needs to be considered and eventually transformed to the relevant level, \eg, this can also involve the user of a system. 
This may be necessary because of a reaction of superimposed systems or the environment due to the behavior of the considered system:
\begin{RQ}[resume]
	\item The fault tolerance regime definitions should be applicable at different system levels.
	\label{req:systemhierarchy}
\end{RQ}

For a complex system like a vehicle, a high number of possible faults can be expected. 
Faults may also occur simultaneously and may lead to different behavior of the system depending on its fault tolerance characteristics as stated, \eg, by~\citet{isermann_2000,isermann_2002}. 
This is addressed by requirement~\ref{req:multifault}:
\begin{RQ}[resume]
	\item The fault tolerance regime definitions must work for an arbitrary number of concurrent faults.%
	\label{req:multifault}
\end{RQ}

We use these requirements as a basis for the development of our definitions and will verify the definitions by these requirements in \autoref{subsec:verification}.

\subsection{Related Terms}
\label{subsec:relatedterms}

In order to define the terms for the fault tolerance regimes in \autoref{subsec:definitions}, it is necessary to provide definitions for related terms.
The related terms are either used directly for the definition of the different fault tolerance regimes or in the explanatory text. %
Most provided related terms are based on ISO\,26262~\cite{iso_2018}. %
When we deviate from the definitions provided by ISO\,26262, we argue our reasons for this deviation.

\begin{term}[Safety]
	Absence of unreasonable risk.~{\normalfont\cite[Part~1, 3.132]{iso_2018}} 
\end{term}
\noindent As a consequence, we assume that there is a risk threshold. 
Below this threshold, the risk is accepted, while above the threshold, the system is considered unsafe.
It is worth noting that we take solely an engineering perspective. 
A legal perspective would potentially demand zero risk~\cite{gasser_2021}.

\begin{term}[Risk]
	Combination of the probability of the occurrence of harm and the severity of that harm.~{\normalfont\cite[Part~1, 3.128]{iso_2018}}
\end{term}%
\noindent Harm in the context of ISO\,26262 always refers to ``physical injury or damage to the health of persons''~\cite[Part~1, 3.74]{iso_2018} and excludes damage to, \eg, property or reputation.
Furthermore, it is our understanding that, when establishing the combination introduced by the definition of \emph{risk}, an increase in either factor, \ie\ the probability of occurrence or the severity, must result in a monotonically increasing risk quantification.

\begin{term}[System]
	An entity that interacts with other entities, \ie, other systems, including hardware, software, humans, and the physical world with its natural phenomena.~{\normalfont\cite[2]{avizienis_2004}}
\end{term}
\noindent According to~\citet[2]{avizienis_2004}, the ``other systems are the environment of the given system. The system boundary is the common frontier between the system and its environment.''
ISO\,26262~\cite[Part~1, 3.163]{iso_2018} defines a system as a ``set of components or subsystems that relates at least a sensor, a controller and an actuator with one another.''
We prefer the definition by~\citeauthor{avizienis_2004} over the definition provided by ISO\,26262 as it provides a broader perspective on the system concept. %
By applying the definition of~\citeauthor{avizienis_2004} to the definition of ISO\,26262, the controller, the sensor, and the actuator as entities can each be considered as a system by themselves with the remaining components belonging to the environment the entity interacts with.
Thus, a narrower system boundary can be drawn compared to ISO\,26262 without excluding the specification of the system boundary of ISO\,26262.

\begin{term}[Fault]
	Abnormal condition that can cause an element or an item to fail.~{\normalfont\cite[Part~1, 3.54]{iso_2018}}
\end{term}
\noindent This definition is in accordance with~\citet{avizienis_2004}, who define a fault as the possible cause of an error, where an error is the system's deviation from its correct external state, \ie, what is perceivable at the system's interfaces to the environment. 
All external states together make up the system's behavior.
While not explicitly stated by ISO\,26262 or~\citeauthor{avizienis_2004}, we assume faults to be discrete and distinguishable, as suggested in Part~5 of ISO\,26262.

\begin{term}[Failure]
	Termination of an intended behavior of an element or an item due to a fault manifestation.~{\normalfont\cite[Part~1, 3.50]{iso_2018}}
\end{term}
\noindent This definition follows the definition by~\citet[3]{avizienis_2004} of a failure as a deviation between the behavior provided by the system and its correct behavior to the point that the system is unable to provide its intended function. 

\begin{term}[Fault tolerance]
	Ability to deliver a specified functionality in the presence of one or more specified faults.~{\normalfont\cite[Part~1, 3.60]{iso_2018}}
\end{term}
\noindent This definition is almost identical to the definition given by \citeauthor{avizienis_2004} for whom fault tolerance ``means to avoid~[...] failures in the presence of faults''~\cite[4]{avizienis_2004}.

\begin{term}[Performance]
	Quantitative measure characterizing a physical or functional attribute relating to the execution of a process, function, activity, or task; performance attributes include quantity (how many or how much), quality (how well), timeliness (how responsive, how frequent), and readiness (when, under which circumstances).~{\normalfont\cite[264]{walden_2015}}
\end{term}
\noindent \emph{Performance} has not been defined so far in automotive safety standards. 
The use of the concept of \emph{performance}, such as in the description of \emph{performance limitations} in the ISO/DIS\,21448 standard, is not accompanied by a clear definition of the term \emph{performance}.
Consequently, we refer to the definition of \emph{performance} from systems engineering.

\subsection{Definitions}
\label{subsec:definitions}
Based on the requirements presented in \autoref{subsec:requirements} together with the related terms in \autoref{subsec:relatedterms}, we propose the following taxonomy for fault tolerance regimes. 
The taxonomy consists of five terms: \op, \failop, \faildegraded, \failsafe, and \failunsafe.
They are supported by a concise definition of \safestate\ as well as by definitions of \emph{functionality}, \emph{available performance}, and \emph{nominal performance}. 
However, the term \emph{fault tolerance regime} must first be defined:
\begin{definition}[{Fault tolerance regime}]
	\label{def:degradationregime}
	System property that classifies the system's behavior in the presence of a specific fault combination. 
\end{definition}
\noindent 
The term \emph{fault combination} is used as distinct faults can occur at the same time.
So, assuming that $\mathcal{F}$ denotes the set of all distinct faults that can occur in a system, $\mathcal{P}(\mathcal{F})=\{f\mid f\subseteq \mathcal{F}\}$ is the set of possible fault combinations~$f$.
The more general assumption of fault combinations used here also covers the ``fail behavior'' in response to an increasing number of faults, which is used by~\citet{jacobson_1996, isermann_2000,isermann_2002}.

Unlike other publications, we argue that fault tolerance regimes are not necessarily a system-wide property.
Rather, fault tolerance regimes are a property that must be seen with respect to a set of covered fault combinations. 
On the one hand, fault tolerance regimes are often connected to an assumption about how many faults can occur at the same time. 
For instance, a single fault assumption $|f|=1$ is regularly used in the automotive domain~\cite{isermann_2000}.
Thus, all fault combinations $\{ f \mid |f|>1\}$ are not covered by the specific fault tolerance regime. 
On the other hand, even within the assumed range of~$|f|$, a single fault combination that is not covered is enough to invalidate a fault tolerance regime assigned to an entire system. 
As a consequence, all fault combinations must be known to be able to assign a specific system-wide fault tolerance regime, which is challenging to say the least.

For the following definitions, we use four criteria to distinguish fault tolerance regimes from each other. 
The \emph{first criterion} is whether a fault is present in the system.
Furthermore, fault tolerance regimes are fundamentally associated with the question of what is safe in the context of a system. 
Thus, the \emph{second criterion} is whether a system allows for a \safestate\ in the presence of a fault combination~$f$. 
We define a \safestate\ as follows:
\begin{definition}[\emph{\Safestate}]
	\label{def:safestate}
	State in which a system does not pose an unreasonable risk.
\end{definition}
With this definition, we generalize inconsistent definitions by ISO\,26262, ISO/DIS\,21448, and ISO/TR\,4804.
A \emph{safe state} according to ISO\,26262 is an ``operating mode, in case of a failure, of an item without an unreasonable level of risk''~\cite[Part~1, 3.131]{iso_2018}.
An \emph{operating mode} refers to the ``conditions of functional state that arise from the use and application of an item or element''~\cite[Part~1, 3.102]{iso_2018}, \eg, ``system off'', ``system active'', ``degraded operation''.

When defining the term \safestate, both ISO/DIS\,21448~\cite{iso_2021} and ISO/TR\,4804~\cite{iso_2020} refer to the \emph{minimal risk condition} introduced by SAE J3016~\cite{sae_2021} but come to different conclusions.
ISO/DIS\,21448 equates the \safestate\ defined by ISO\,26262 with a \emph{minimal risk condition}, which is a vehicle state that is supposed ``to reduce the risk of harm, when a given trip cannot be completed''~\cite[\nopp 3.16]{iso_2021}.
In contrast, ISO/TR\,4804 distinguishes between \safestate\ and \emph{minimal risk condition} and defines a \safestate\ as an ``operating mode that is reasonably safe''~\cite[\nopp 3.50]{iso_2020}, while a \emph{minimal risk condition} is a ``condition to which a user or an automated driving system may bring a vehicle after performing the \emph{minimal risk manoeuvre} in order to reduce the risk of a crash when a given trip cannot be completed''~\cite[\nopp 3.29]{iso_2020}. 
Consequently,  a \emph{minimal risk condition} is not necessarily safe. 
Instead, it is the condition a vehicle will reach in response to specific events due to the implemented safety mechanisms. 
This condition may exceed the accepted risk threshold and would therefore not qualify as a \safestate.
Furthermore, the understanding of \safestate\ in ISO/TR\,4804~\cite[\nopp 3.50]{iso_2020} reflects the understanding of \safestate\ outlined by~\citet{reschka_2015c}.
\citeauthor{reschka_2015c} argue at the vehicle level that an (automated) vehicle must maintain a \safestate\ even without the occurrence of a fault.

Although the definition of the \safestate\ provided by ISO/TR\,4804 appears reasonably generic, we provide a more general definition for the term \safestate\ in \autoref{def:safestate} for two reasons.
Firstly, the definition by ISO/TR\,4804 is only intended to be applied at the vehicle level and not at a subsystem level.
Secondly, it relies on the term \emph{operating mode}, which however is not further explained in ISO/TR\,4804. 
It presumes knowledge of the corresponding definition in ISO\,26262~\cite[Part 1, 3.131]{iso_2018}, though this definition and the associated examples are inconclusive.

Our understanding of the term \safestate\ according to \autoref{def:safestate} is in principle applicable to all kinds of safety considerations: those targeting internal faults as well as those targeting insufficient specification or unconsidered technological limitations. 
However, the following definitions of fault tolerance regimes presume a sufficient system specification as well as a complete consideration of technological limitations. 
In the context of E/E automotive systems, our focus is on \emph{functional safety} according to ISO\,26262~\cite{iso_2018} rather than on the \emph{safety of the intended functionality} according to ISO/DIS\,21448~\cite{iso_2021}.
Still, the definition of \safestate\ works for both.
Moreover, arguing that a system does not pose an unreasonable risk often requires consideration of neighboring or superimposed systems because a \safestate\ preservation could require an adequate reaction of these.

For describing a system, we distinguish between the system's \emph{functionality} and its \emph{performance} while providing the \emph{functionality}.
This distinction follows \citet{avizienis_2004} who propose that a system is specified by a dualism of \emph{functionality} and \emph{performance}. 
Still, neither term is defined in~\cite{avizienis_2004} or in automotive E/E safety standards.
Yet, a definition of \emph{performance} is given in systems engineering context~\cite{walden_2015}, \cf\ \autoref{subsec:relatedterms}. 
For \emph{functionality}, we propose the following definition:
\begin{definition}[\emph{Functionality}]
	Behavior of a system expressed in its interaction with its operating environment.
\end{definition}
\noindent This definition integrates the term \emph{behavior}, which is frequently encountered in ISO\,26262 and ISO/DIS\,21448 to describe what a system does, with a description of functionality by \citet{walden_2015}.
\citeauthor{walden_2015} state that ``the functionality of a system is typically expressed in terms of the interaction of the system with its operating environment [...]'' \cite[6]{walden_2015}.%

Describing a system by means of \emph{functionality} and \emph{performance}, allows establishing the third and fourth criterion for the distinction of fault tolerance regimes.
The \emph{third criterion} is the system's ability to provide its functionality. 
In general, a \safestate\ can be maintained both when a system provides its functionality and when it does not. 
Therefore, we integrate the availability of a \safestate\ with the system's ability to provide its specified functionality into the system's operability $o(f)$ as \newline
\begin{equation*}
	o(f) = \begin{cases}
		\phantom{-}1, & \parbox[t]{6.25cm}{system is in a \safestate\ while providing its specified functionality;}\\
		\phantom{-}0, & \parbox[t]{6.25cm}{system is in a \safestate\ while not providing its specified functionality;}\\
		- 1,		  & \parbox[t]{6.25cm}{otherwise.}
	\end{cases}	
\end{equation*}

The \emph{fourth} and \emph{last criterion} used for distinguishing \emph{fault tolerance regimes} is the \emph{available performance}~$p_\mathrm{a}(f)$ of the system while providing its functionality:
\begin{definition}[Available performance]
	Performance that is available for a system to provide its specified functionality. 
\end{definition}

In order to allow for a distinction of fault tolerance regimes, the \emph{available performance} can be related to the \emph{nominal performance}~$p_\mathrm{nom}$, which we define as follows: 
\begin{definition}[Nominal performance]
	Performance with which a fault-free system is expected to be able to provide its specified functionality.
\end{definition}

\begin{figure}%
	\centering
	\setlength{\nodeheight}{\heightof{\footnotesize fail-operational}}
\setlength{\nodeheight}{1.1\nodeheight}
\newlength{\nodedepth}
\setlength{\nodedepth}{\depthof{\footnotesize fail-operational}}
\setlength{\nodedepth}{1.1\nodedepth}

\newlength{\nodeVdistance}
\setlength{\nodeVdistance}{\totalheightof{\footnotesize yes}}
\setlength{\nodeVdistance}{1.5\nodeVdistance}
\newlength{\nodeHdistance}
\setlength{\nodeHdistance}{\widthof{\footnotesize yes}}
\setlength{\nodeHdistance}{1.2\nodeHdistance}

\newlength{\nodewidth}
\setlength{\nodewidth}{\linewidth}
\addtolength{\nodewidth}{-3\nodeHdistance}
\setlength{\nodewidth}{\nodewidth/4}
\addtolength{\nodeHdistance}{\nodewidth}

\tikzset{
	treenode/.style = {
		shape=rectangle, rounded corners,
		draw, align=center,
		top color=white, bottom color=ourBlue,
		inner sep=0,
		font=\footnotesize,
	},
	criterion/.style    = {treenode, bottom color=ourRed},
	regime/.style      	= {treenode, anchor=base,text height=\nodeheight,text depth=\nodedepth}
}
\begin{tikzpicture}[node distance=3em and \nodewidth,font=\footnotesize]

	\newcommand{\drawblock}[5][draw]{
		\node[fit=(#2)(#3),#1] (#4) {};
		\node[align=center,minimum height=\nodeheight/1.1]at (#4.center)(){#5};
	}

	\matrix (m) [	
		matrix of nodes,
		nodes={
			align=center, 
			text width=\nodewidth,
			inner sep=0,
			anchor=center,
		},
		nodes in empty cells,
		column sep={\nodeHdistance,between origins},%
		row sep={\nodeVdistance,between borders},%
		row 1/.style={nodes={minimum height=.45cm}},
		row 2/.style={nodes={minimum height=1.25cm}},
		row 3/.style={nodes={minimum height=.45cm}},
	]
	{
		&&&|[regime](failop)|fail-operational\\
		&&&\\	
		|[regime](operational)|operational&
		|[regime](failunsafe)|fail-unsafe&
		|[regime](failsafe)|fail-safe&
		|[regime](faildegraded)|fail-degraded\\
	};
	\node[circle, fill=black] at (m-1-1) (start) {};
	\drawblock[criterion]{m-2-1.north west}{m-2-1.south east}{fault}{Fault present\\in system?}
	\drawblock[criterion]{m-2-2.north west}{m-2-2.south east}{safestate} {System is\\intentionally\\maintaining\\safe state?};
	\drawblock[criterion]{m-2-3.north west}{m-2-3.south east}{functional} {System is\\providing its\\functionality?};
	\drawblock[criterion]{m-2-4.north west}{m-2-4.south east}{performance} {Available\\performance\\$\geq$ nominal\\performance?};

	\draw[-latex] (start) 			-- (fault);
	\draw[-latex] (fault) 			-- node[midway, right,anchor=base west,yshift=-.05cm] {no} (operational);
	\draw[-latex] (safestate) 		-- node[midway, right,anchor=base west,yshift=-.05cm] {no} (failunsafe);
	\draw[-latex] (functional) 		-- node[midway, right,anchor=base west,yshift=-.05cm] {no} (failsafe);
	\draw[-latex] (performance) 	-- node[midway, right,anchor=base west,yshift=-.05cm] {yes} (failop);
	\draw[-latex] (performance) 	-- node[midway, right,anchor=base west,yshift=-.05cm] {no} (faildegraded);
	
	\draw[-latex] (fault) 			-- node[midway, above,anchor=base,yshift=.1cm] {yes} (safestate);
	\draw[-latex] (safestate)		-- node[midway, above,anchor=base,yshift=.1cm] {yes} (functional);
	\draw[-latex] (functional) 		-- node[midway, above,anchor=base,yshift=.1cm] {yes} (performance);

	\pgfresetboundingbox
	\path [use as bounding box,inner sep=0, line width=0,] (operational.south west) rectangle (failop.north east);
	
\end{tikzpicture}
	\caption{Scheme to distinguish fault tolerance regimes according to the presented taxonomy.}
	\label{fig:tree}
\end{figure}
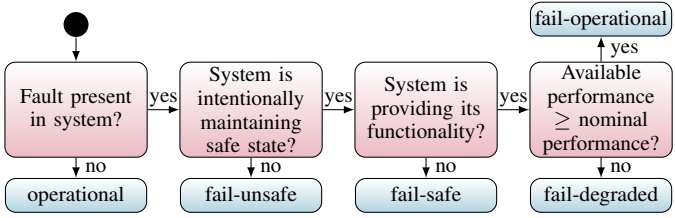

As illustrated by the scheme presented in \autoref{fig:tree}, the evaluation of the four criteria in the order of their appearance in this section allows for a clear distinction of fault tolerance regimes.
Thus, it facilitates their definitions, which are outlined in the following paragraphs.
As our focus is on fault tolerance, we presume as a starting point that a fault-free system is always able to maintain a \safestate.
\begin{definition}[\emph{\Op}]
	\label{def:op}
	An \emph{\op}\ system has no fault and, thus, can provide its specified functionality with at least nominal performance while maintaining a safe state. 
\end{definition}
\noindent Therefore, a system is \op\ for $f=\emptyset \subseteq \mathcal{P}(\mathcal{F})$, from which follows $o(\emptyset)=1$ and $p_\mathrm{a}(\emptyset)\geq p_\mathrm{nom}$.

\begin{definition}[\emph{\Failunsafe}]
	\label{def:failunsafe}
	A system is \emph{\failunsafe}\ in the presence of a fault combination if it is not able to maintain a safe state.
\end{definition}
\noindent Consequently, $\{f \mid o(f)=-1\} \subseteq \mathcal{P}(\mathcal{F})$ defines the set of fault combinations that a system cannot handle safely.

\begin{definition}[\emph{\Failsafe}]
	\label{def:failsafe}
	A system is \emph{\failsafe}\ in the presence of a  fault combination if it ceases its specified functionality and transitions to a well-defined condition to maintain a safe state. 
\end{definition}
\noindent Thus, $\{f \mid o(f)=0\} \subseteq \mathcal{P}(\mathcal{F})$ defines the set of fault combinations that a system can handle in a \failsafe\ manner.
It is important to note again that whether a \safestate\ is suitable can only be assessed by considering neighboring or superimposed systems. 
For instance, the \emph{minimal risk condition} described for automated vehicles in ISO/TR\,4804 and ISO/DIS\,21448 is a potential \safestate. 
Other road users, who may be considered neighboring systems, may have to adapt to an automated vehicle that, \eg, pulls over to the side of the road after a fault.

We do not use the term \failsilent\ in this taxonomy for two reasons. 
Firstly, \failsafe\ and \failsilent\ are not sharply differentiated in literature as shown in \autoref{sec:stateoftheart}. 
Similar to the \safestate, we understand a ``silent'' system output as a specific defined state. 
Secondly, \failsilent\ as a property that requires systems to have no output at all is hard to imagine. 
For systems that are required to be continuously active, a suddenly inactive output carries information as well. 
In contrast, for systems that are only rarely used, a system shutdown indication is common. 
For example, Electronic Stability Control systems, \cf~\cite{isermann_2000,isermann_2002}, usually feature an indication to the driver if the systems have encountered a shutdown.

\begin{definition}[\emph{\Faildegraded}]
	\label{def:faildegraded}
	A system is \emph{\faildegraded}\ in the presence of a fault combination if it can provide its specified functionality with below nominal performance while maintaining a safe state. 
\end{definition}
\noindent Thus, $\{f \mid (o(f)=1) \wedge (p_\mathrm{a}(f)<p_\mathrm{nom})\}\subseteq \mathcal{P}(\mathcal{F})$ defines the set of fault combinations that a system can handle in a \faildegraded\ manner.
In contrast to \autoref{def:op}, the available performance~$p_\mathrm{a}(f)$ is lower than the nominal performance~$p_\mathrm{nom}$ while the overall \safestate\ is maintained. 
\Faildegraded\ is chosen rather than \failreduced\ because it is more common in the body of literature we have reviewed.
An operation with an available performance~$p_\mathrm{a}$ below the nominal performance~$p_\mathrm{nom}$ may require an adaptation to this degradation on superimposed system layers. %
An example for a \faildegraded\ behavior is the limp home mode of combustion engines, which allows for a continuation of a trip with reduced speed in the presence of a fault~\cite{schauffele_2016, chen_2008, isermann_2002, schnellbach_2016a}. 

\begin{definition}[\emph{\Failop}]
	\label{def:failop}
	A system is \emph{\failop}\ in the presence of a fault combination if it can provide its specified functionality with at least nominal performance while maintaining a safe state. 
\end{definition}
\noindent Consequently, $\{f \mid (o(f)=1) \wedge (p_\mathrm{a}(f)\geq p_\mathrm{nom})\}\subseteq \mathcal{P}(\mathcal{F})$ defines the set of fault combinations that a system can handle in \failop\ manner.
Contrary to some researchers' understanding of \failop, we do not subsume a degraded operation in \failop. 
As for \autoref{def:op}, the available performance~$p_\mathrm{a}(f)$ equals at least nominal performance~$p_\mathrm{nom}$ while the overall \safestate\ is maintained.
Thus, neighboring or superimposed systems can continue their operation as with an \op\ system.

\section{APPLICATION AND VERIFICATION OF THE PROPOSED TAXONOMY}
\label{sec:application}
To demonstrate that this taxonomy can be applied to different systems as well as at different system levels, we introduce two examples from the automotive domain. 
\autoref{subsec:steerbywire} illustrates the taxonomy at the example of a steer-by-wire system.
As a more complex example, we show in \autoref{subsec:safehalt} that the taxonomy can be applied to an automated driving application as well. 
Both examples contribute to the verification of the taxonomy against the requirements, which is presented in \autoref{subsec:verification}.

\subsection{Steer-by-Wire System}
\label{subsec:steerbywire}
A steer-by-wire system is the first example for evaluating the taxonomy. 
Steer-by-wire is considered as highly safety-critical in general and is a necessary feature of future automated vehicles. 
In the steer-by-wire system, the desired steering angle is calculated and controlled purely by an electronic system. 

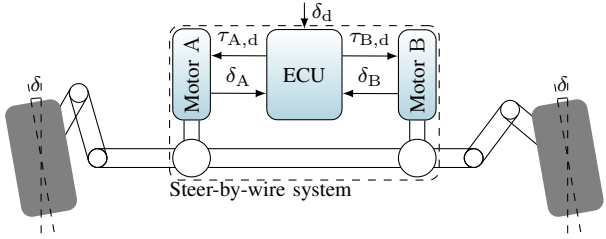
\begin{figure}
	\centering
	\begin{tikzpicture}
	\footnotesize
	\coordinate (wheelleft)at (-3.5cm,0.0cm){};
	\coordinate (wheelright)at ($(wheelleft.center)+(7cm,-0.0cm)$){};
	\coordinate (jointleft)at ($(wheelleft.center)+(60:1cm)$);
	\coordinate (jointright)at ($(wheelright.center)+(140:1cm)$);
	\coordinate (steeringrackleft)at (-2.75cm,0);;
	\coordinate (steeringrackright) at ($(steeringrackleft)+(5cm,0)$);
	
	\node [circle, minimum size=.25cm,draw,inner sep =0] at (jointleft) (){};
	\node [circle, minimum size=.25cm,draw,inner sep =0] at (jointright) (){};
	\draw []($(wheelleft)+(150:0.125cm)$) to ($(jointleft)+(150:.125cm)$);
	\draw []($(wheelleft)+(-30:0.125cm)$) to ($(jointleft)+(-30:.125cm)$);
	\draw []($(wheelright)+( 50:0.125cm)$) to ($(jointright)+( 50:.125cm)$);
	\draw []($(wheelright)+(230:0.125cm)$) to ($(jointright)+(230:.125cm)$);
	\node [fill=gray, rounded corners, minimum height=1.5cm,minimum width=.75cm,inner sep=0,rotate=10,anchor=center] at (wheelleft)(){};
	\node [fill=gray, rounded corners, minimum height=1.5cm,minimum width=.75cm,inner sep=0,rotate=10,anchor=center] at (wheelright)(){};
	\draw [dashed] ($(wheelleft)+(-90:1cm)$) to ($(wheelleft)+(90:1cm)$);
	\draw [dashed] ($(wheelleft)+(-80:1cm)$) to ($(wheelleft)+(100:1cm)$);
	\draw [dashed] ($(wheelright)+(-90:1cm)$) to ($(wheelright)+(90:1cm)$);
	\draw [dashed] ($(wheelright)+(-80:1cm)$) to ($(wheelright)+(100:1cm)$);
	\draw [] ($(wheelleft)+(90:.8cm)$) arc (90:100:0.8cm);
	\draw [] ($(wheelright)+(90:.8cm)$) arc (90:100:0.8cm);
	\node [anchor=south] at ($(wheelleft)+(95:.8cm)$)(){$\delta$};
	\node [anchor=south] at ($(wheelright)+(95:.8cm)$)(){$\delta$};

	\draw []($(steeringrackleft)+(0,0.125cm)$) to ($(steeringrackright)+(0,0.125cm)$);
	\draw []($(steeringrackleft)-(0,0.125cm)$) to ($(steeringrackright)-(0,0.125cm)$);
	\node[circle, minimum size=.25cm,draw,inner sep =0] at (steeringrackleft)(steeringrackleft){};
	\node[circle, minimum size=.25cm,draw,inner sep =0] at (steeringrackright)(steeringrackright){};

 	\pgfmathanglebetweenpoints{\pgfpointanchor{jointleft}{center}}{\pgfpointanchor{steeringrackleft}{center}}
	\let\angle\pgfmathresult
	\path let \p1=(jointleft), \p2=(steeringrackleft), \n1={veclen(\x2-\x1,\y2-\y1)} in node[outer sep=0,inner sep=0, minimum height = 0.25cm, rotate=\angle, minimum width=\n1,anchor=west] at (jointleft) (test){};
	\draw (test.north west) to (test.north east);
	\draw (test.south west) to (test.south east);
	
 	\pgfmathanglebetweenpoints{\pgfpointanchor{jointright}{center}}{\pgfpointanchor{steeringrackright}{center}}
	\let\angle\pgfmathresult
	\path let \p1=(jointright), \p2=(steeringrackright), \n1={veclen(\x2-\x1,\y2-\y1)} in node[outer sep=0,inner sep=0, minimum height = 0.25cm, rotate=\angle, minimum width=\n1,anchor=west] at (jointright) (test){};
	\draw (test.north west) to (test.north east);
	\draw (test.south west) to (test.south east);
		
	\coordinate (pinionleft) at (-1.5cm,0){};
	\coordinate (pinionright) at ( 1.5cm,0){};
	\coordinate (sensor) at ( 0,0){};
	\node [draw, rounded corners,anchor=west, rotate=90,minimum width=1.2cm, minimum height=.5cm,top color=white, bottom color=ourBlue] at ($(pinionleft)+(0,.5cm)$)(motora){Motor $\mathrm{A}$};
	\node [draw, rounded corners,anchor=west, rotate=90,minimum width=1.2cm, minimum height=.5cm,top color=white, bottom color=ourBlue] at ($(pinionright)+(0,.5cm)$)(motorb){Motor $\mathrm{B}$};
	\node [draw, rounded corners,anchor=north west,inner sep=0,align=center, minimum height=1.2cm,minimum width=1cm,anchor=center,top color=white, bottom color=ourBlue] at ($(motora)!0.5!(motorb)$)(ecu){ECU};

	\draw [] ($(motora.west)+(-0.1cm,0)$) to ($(pinionleft)+(-0.1cm,0)$);
	\draw [] ($(motora.west)+( 0.1cm,0)$) to ($(pinionleft)+( 0.1cm,0)$);
	\draw [] ($(motorb.west)+(-0.1cm,0)$) to ($(pinionright)+(-0.1cm,0)$);
	\draw [] ($(motorb.west)+( 0.1cm,0)$) to ($(pinionright)+( 0.1cm,0)$);
	\node [circle, minimum size=.5cm,draw,fill=white]at (pinionleft)(pinionleft) {};
	\node [circle, minimum size=.5cm,draw,fill=white]at (pinionright)(pinionright) {};

	\node[circle, minimum size=.25cm,draw,inner sep =0] at (steeringrackleft)(tierodleft){};
	\node[circle, minimum size=.25cm,draw,inner sep =0] at (steeringrackright)(tiedright){};

	\draw [-latex] ($(ecu.north)+(0,0.35cm)$) node (topnode){} to (ecu.north) ;
	\node [anchor=south west] at (ecu.north){$\delta_\mathrm{d}$}; 
	
	\draw [-latex] ($(ecu.west)+(0,.25cm)$) to node[midway,above](){$\tau_\mathrm{A,d}$} ($(motora.south)+(0,.25cm)$);
	\draw [-latex] ($(ecu.east)+(0,.25cm)$) to node[midway,above](){$\tau_\mathrm{B,d}$} ($(motorb.north)+(0,.25cm)$);
	\draw [latex-] ($(ecu.west)-(0,.25cm)$) to node[midway,above](){$\delta_\mathrm{A}$} ($(motora.south)-(0,.25cm)$);
	\draw [latex-] ($(ecu.east)-(0,.25cm)$) to node[midway,above](){$\delta_\mathrm{B}$} ($(motorb.north)-(0,.25cm)$);

	\node [draw, dashed,rounded corners,fit=(motora.north east)(pinionright.south-|motorb.south),anchor=center,inner sep=1pt] at ($0.5*(motora.north east)+0.5*(pinionright.south-|motorb.south)$)(system){};
	\node [anchor=north west,inner sep=0,align=left] at (system.south west)(){Steer-by-wire system};
\end{tikzpicture}
	\caption{Sketch of steer-by-wire system. 
		ECU:~electronic control unit; 
		$\delta$:~steering angle;
		$\delta_\mathrm{d}$:~desired steering angle;
		$\delta_\mathrm{A}$:~motor angle of motor $\mathrm{A}$;
		$\delta_\mathrm{B}$:~motor angle of motor $\mathrm{B}$;
		$\tau_{\mathrm{A,d}}$:~desired torque of motor $\mathrm{A}$;
		$\tau_{\mathrm{B,d}}$:~desired torque of motor $\mathrm{B}$.%
	}
	\label{fig:SbW}
\end{figure}

\autoref{fig:SbW} illustrates the example steer-by-wire system used in this subsection, which consists of three subsystems.
There are the two steering motors~$\mathrm{A}$ and~$\mathrm{B}$.
The two motors apply torques~$\tau_{\mathrm{A}}$ and~$\tau_{\mathrm{B}}$ to the steering rack in order to control the steering angle~$\delta$ by shifting the steering rack to the left or right. 
They each can provide motor angles~$\delta_\mathrm{A}$ and~$\delta_\mathrm{B}$, which directly relate to the actual steering angle~$\delta$. 
The third subsystem is the electronic control unit (ECU), which closes the steering angle control feedback loop. %

From a functional perspective, the system is required to steer in terms of \emph{setting the steering angle}.
Thus, the actual steering angle~$\delta$ is the system's output while the desired steering angle~$\delta_\mathrm{d}$ is the input.
Internally, the ECU calculates a required steering torque~$\tau_\mathrm{d}$ and allocates it to motors~$\mathrm{A}$ and~$\mathrm{B}$.
They are subject to~$\tau_\mathrm{d}=\tau_{\mathrm{A},\mathrm{d}}+\tau_{\mathrm{B},\mathrm{d}}$, where~$\tau_{\mathrm{A},\mathrm{d}}$ and~$\tau_{\mathrm{B},\mathrm{d}}$ are the desired torques for motor~$\mathrm{A}$ and~$\mathrm{B}$.
Finally, the resulting steering torque $\tau=\tau_\mathrm{A}+\tau_\mathrm{B}$ alters the actual steering angle~$\delta$.

Let's assume that the target application requires a minimum steering angle range~$\delta_\mathrm{nom}\in [\ubar{\delta}~\bar{\delta}]$ 
as well as minimal available steering torque range~$\tau_\mathrm{nom}\in [\ubar{\tau}~\bar{\tau}]$ where $\bar{\cdot}$ and $\ubar{\cdot}$ indicate the upper and lower bounds of the ranges. 
Summarized, this yields the steer-by-wire system's nominal performance~$p_\mathrm{SbW,nom}= p_\mathrm{SbW}(\delta_\mathrm{nom},\tau_\mathrm{nom})$.

Similarly, the nominal performances of the subsystems can be defined.
For motors~$\mathrm{A}$ and~$\mathrm{B}$,
these are $p_\mathrm{A,nom}= p_\mathrm{A}(\tau_\mathrm{A,nom})$ with $\tau_\mathrm{A,nom}\in [\ubar{\tau}_\mathrm{A}~\bar{\tau}_\mathrm{A}]$ and $p_\mathrm{B,nom}=p_\mathrm{B}(\tau_\mathrm{B,nom})$ with $\tau_\mathrm{B,nom} \in [\ubar{\tau}_\mathrm{B}~\bar{\tau}_\mathrm{B}]$. 
For the ECU, a performance measure is, \eg, the number of operations per second. 
Further explanations regarding the ECU subsystem are omitted as we consider it as \failop\ for this example. 

Let's assume that motor~$\mathrm{A}$ is subject to a fault combination~$f_\mathrm{A}$ and that the motor comes with a mechanism targeting \failsafe\ behavior for~$f_\mathrm{A}$. 
The mechanism forces motor~$\mathrm{A}$ into a torque-free state~($\tau_\mathrm{A,fs}=0$) and, thus, inhibits its ability to steer while the motor maintains a defined state.
Assessing whether motor~$\mathrm{A}$ with zero steering torque still results in a safe steer-by-wire system requires taking the remaining system components into account, in particular motor~$\mathrm{B}$. 
If $\tau_\mathrm{B,nom}\supseteq\tau_\mathrm{nom}$, the steer-by-wire system is \failop\ in case of the fault combination~$f_\mathrm{A}$.
$\tau_\mathrm{B,nom}\subset\tau_\mathrm{nom}$ yields a \faildegraded\ system as the steering is still functional, $o_\mathrm{SbW}(f_\mathrm{A})=1$, yet with decreased available performance. 
For the latter, $o_\mathrm{SbW}(f_\mathrm{A})=1$ presumes that the resulting available performance~$p_\mathrm{SbW,a}(f_\mathrm{A})=p_\mathrm{SbW}(\delta_\mathrm{a}(f_\mathrm{A}),\tau_\mathrm{a}(f_\mathrm{A}))$ suffices to operate the vehicle safely although it is below its nominal performance, $p_\mathrm{SbW,a}(f_\mathrm{A})<p_\mathrm{SbW,nom}$. 
It follows $o_\mathrm{A}(f_\mathrm{A})=0$ such that motor~$\mathrm{A}$ is \failsafe\ for $f_\mathrm{A}$.

Altogether, this example illustrates two different aspects of the taxonomy.
Firstly, it demonstrates that the taxonomy can be applied to hierarchical system designs. 
Secondly, it also shows that the nominal performance is not necessarily congruent to the maximum available performance. 
Again, it is important to note that the fault tolerance regimes are always related to a specific fault combination.

\subsection{SAE Level 4 Automated Driving Application}
\label{subsec:safehalt}

\begin{figure}[b]
	\centering
	\vspace{1em}
	\begin{tikzpicture}
	\footnotesize
	\newcommand{\drawblock}[5][draw]{
		\node[fit=(#2)(#3),box,#1] (#4) {};
		\node[align=center]at (#4)(){#5};
	}

	\tikzstyle{box} = [rounded corners,inner sep=0,outer sep=0,align=center,text centered,minimum width=1.5cm]
	\matrix (m) [	matrix of nodes,
				 	nodes={
				 		align=center, 
				 		text width=1.5cm,
				 		inner sep=0,
				 		anchor=center,
			 		},
		 			column sep=.35cm,%
	 				row sep=.35cm,%
		 			nodes in empty cells,
		 			row 1/.style={nodes={minimum height=.5cm}},
		 			row 2/.style={nodes={minimum height=1.0cm}},
		 			row 3/.style={nodes={minimum height=.5cm}},
		 			row 4/.style={nodes={minimum height=.75cm}},
				]
	{
		 &&&\\
		 &&&\\	
		 &&&\\
		 &&&\\
	};

	\draw [rounded corners,top color=white, bottom color=ourBlue] 
		(m-1-1.north west)
		--(m-1-4.north east)
		--(m-1-4.south east)
		--(m-1-1.south east)
		--(m-4-1.south east)
		--(m-4-1.south west)
		--cycle; 
	\draw [rounded corners,dashed]
		($(m-1-1.north west)+(-2pt,2pt)$)
		--($(m-1-4.north east)+(2pt,2pt)$)
		--($(m-4-4.south east)+(+2pt,-2pt)$)
		--($(m-4-1.south west)+(-2pt,-2pt)$)
		--cycle;

	\draw [rounded corners,top color=white, bottom color=ourBlue]
		($(m-2-2.north west)$)
		--($(m-2-3.north east)$)
		--($(m-2-3.south east)$)
		--($(m-2-3.south east-|m-4-2.south east)$)
		--($(m-4-2.south east)$)
		--($(m-4-2.south west)$)
		--cycle;

	\drawblock[]{m-1-3.north west}{m-1-4.south east}{TrajGen}{}
	\drawblock[]{m-1-1.north west}{m-4-1.south east}{ADstack}{Normal\\operation\\automated\\driving\\functionality}

	\drawblock[]{m-2-3.north west}{m-2-3.south east}{SiATrajGen}{Safe Halt\\trajectory\\generation}
	\drawblock[]{m-2-2.north west}{m-4-2.south east}{SiAPerception}{Independent\\environment\\perception\\\&\\obstacle\\detection}
	\drawblock[top color=white, bottom color=ourBlue,draw]{m-3-3.north west}{m-3-4.south east}{selector}{Trajectory selection}
	\drawblock[top color=white, bottom color=ourBlue,draw]{m-4-3.north west}{m-4-4.south east}{control}{Vehicle motion control\\and actuation}

	\draw [rounded corners, gray!50] ($(SiAPerception.north west)+(+1pt,-1pt)$)--($(SiAPerception.north east)+(-1pt,-1pt)$)--($(SiAPerception.south east)+(-1pt,+1pt)$)--($(SiAPerception.south west)+(+1pt,+1pt)$)--cycle;
	\draw [rounded corners, gray!50] ($(SiATrajGen.north west)+(+1pt,-1pt)$)--($(SiATrajGen.north east)+(-1pt,-1pt)$)--($(SiATrajGen.south east)+(-1pt,+1pt)$)--($(SiATrajGen.south west)+(+1pt,+1pt)$)--cycle;
	
	\node [anchor=south west,inner sep=0,align=left] at ($(ADstack.north west)+(-2pt,+3pt)$)(){Automated driving system};
	\node [anchor=south west,inner sep=0,align=left] at ($(SiAPerception.north west)+(0pt,1pt)$)(){Safe Halt};
	\draw [-latex] (TrajGen.south-|SiATrajGen) 						to node [midway,right](){\circled{3}}	(SiATrajGen);
	\draw [-latex] (TrajGen.south-|m-2-4) 							to node [midway,right](){\circled{2}}	(m-2-4|-selector.north);
	\draw [-latex] (SiATrajGen.south)		 						to node [midway,right](){\circled{5}}	(SiATrajGen|-selector.north);
	\draw [-latex] (SiAPerception.east|-SiATrajGen) 				to node [midway,above](){\circled{4}}	(SiATrajGen);
	\draw [-latex] (selector) 										to node [midway,right](){\circled{6}}	(control);	
	\draw [-latex] ($.5*(m-1-2.north)+.5*(m-1-3.north)+(0,.35cm)$)	to node [midway,right](){\circled{1}}	($.5*(m-1-2.north)+.5*(m-1-3.north)$);		
	\draw [-latex] (control.south)										to node [midway,right](){\circled{7}}	++(0,-0.35cm);

\end{tikzpicture}
	\caption{
		Functional sketch of an automated driving system with \emph{Safe Halt} emergency stop application. 
		The illustration is based on the generic functional system architecture presented by \citet{matthaei_2015,ulbrich_2017a}.
		\protect\circled{1}~Desired mission;
		\protect\circled{2}~Normal reference trajectory;
		\protect\circled{3}~Emergency path with maximum speed profile along the path;
		\protect\circled{4}~Obstacle list;
		\protect\circled{5}~Emergency reference trajectory;
		\protect\circled{6}~Selected reference trajectory;
		\protect\circled{7}~Vehicle behavior. 
	}
	\label{fig:sia}
\end{figure}
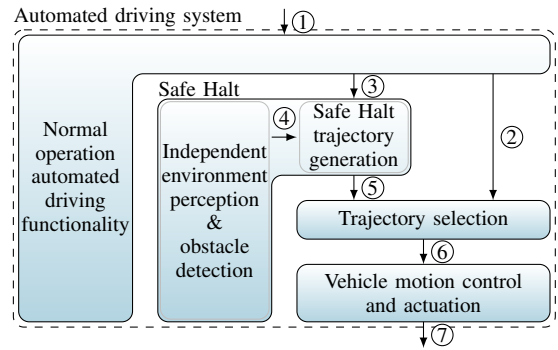

As a second example, we apply the taxonomy to an automated vehicle equipped with an SAE level 4 automated driving system (ADS) according to SAE~J3016~\cite{sae_2021}. 
This ADS determines the vehicle behavior and is therefore highly safety-critical at vehicle level.
\autoref{fig:sia} illustrates the example ADS used here, which consists of four subsystems: the normal operation automated driving functionality, the emergency stop system \emph{Safe Halt}, the trajectory selection, and the vehicle motion control and actuation subsystem.
From a functional perspective, the ADS is required to realize a desired mission~\circled{1}, which is the system's input. 
As output~\circled{7}, the ADS generates the vehicle behavior necessary for mission accomplishment. 
Internally, the normal operation automated driving functionality, which includes the environment perception and interpretation as well as the behavior planning and trajectory generation, outputs a reference trajectory~\circled{2}, which is used for normal vehicle operation. 

As second output, the subsystem generates an emergency path with associated maximum speed profile~\circled{3}.
This emergency path and speed profile are the inputs of the \emph{Safe Halt} functionality~\cite{ackermann_2020}, which is developed in the German research project UNICAR\textit{agil}~\cite{woopen_2018}. 
The emergency path and maximum speed profile are supposed to transition the vehicle from the current vehicle state to a defined state that qualifies as the minimal risk condition for the vehicle according to ISO/DIS\,21448~\cite{iso_2021}. 
Thus, the normal operation automated driving functionality determines the minimal risk condition and plans the emergency path in combination with a path-related maximum speed profile to implement a minimal risk maneuver~\cite{iso_2021} that leads to the minimal risk condition.

The emergency stop system \emph{Safe Halt} determines the emergency reference trajectory~\circled{5} based on the reference path, the maximum speed profile along the path~\circled{3}, and an obstacle list~\circled{4}. 
The obstacle list is generated by an independent environment perception and obstacle detection system, which monitors the driving corridor along the emergency path to avoid collisions with obstacles.
Consequently, the provided obstacle list enables planning a collision-free velocity profile for the minimal risk maneuver.
Based on the emergency path and the collision-free velocity profile, the \emph{Safe Halt} trajectory generation provides the emergency reference trajectories~\circled{5}.

The third subsystem is a trajectory selection that selects the reference trajectory~\circled{6} to be forwarded to the vehicle motion control and actuation.
Based on the health status of the ADS subsystems, either the normal reference trajectory or the emergency reference trajectory is selected.
Finally, the vehicle motion control and actuation subsystem realizes the vehicle behavior based on the selected reference trajectory.

For describing the performance of the ADS, we assume that the ADS is designed to offer a set of selectable missions~$\mathcal{M}_{\mathrm{nom}}$ within its operational design domain~(ODD, \cf~\cite{sae_2021,koopman_2019,bsi_2020} for explanations). %
Furthermore, for each mission~$m$, a mission quality~$q_\mathrm{nom}$ shall be required so that $\forall~m\in\mathcal{M}_\mathrm{nom}~\exists~q_\mathrm{nom}(m)$, which could contain, \ia, the mission execution time as well as measures for driving comfort.

With~$\vec{q}_\mathrm{m,nom}$ containing all $q_\mathrm{nom}(m)$, $p_\mathrm{ADS,nom}=p_\mathrm{ADS}(n_\mathrm{nom},\vec{q}_\mathrm{m,nom})$ is the nominal performance of the ADS, where ${n}_{\mathrm{nom}}=|{\mathcal{M}}_{\mathrm{nom}}|$ denotes the number of nominally selectable missions.
Consequently, the available performance~$p_\mathrm{ADS,a}(f)$ in the presence of a fault combination~$f$ can be described as $p_\mathrm{ADS,a}(f)~=~p_\mathrm{ADS}(n_\mathrm{a}(f),\vec{q}_\mathrm{m,a}(f))$, where $\vec{q}_\mathrm{m,a}(f)$ contains the achievable mission quality in the presence of the fault combination~$f$ $\forall m\in \mathcal{M}_\mathrm{nom}$.

Let's assume that the normal operation automated driving functionality (NADF) is subject to a fault combination~$f_\mathrm{NADF}$.
This fault combination~$f_\mathrm{NADF}$ can affect both performance measures, the number of available missions as well as the achievable mission quality.
If $p_\mathrm{ADS,a}=p_\mathrm{NADF,a}=p_\mathrm{ADS}(f_\mathrm{NADF})\geq p_{\mathrm{NADF,nom}}$, the normal operation automated driving functionality and, thus, the automated driving system is \failop\ for the fault combination~$f_\mathrm{NADF}$. 
This means that the ADS can perform all specified missions in its ODD with the nominal mission quality, yielding $n_\mathrm{a}(f_\mathrm{NADF})=n_\mathrm{nom}$ and $\vec{q}_\mathrm{m,a}(f_\mathrm{NADF})\geq\vec{q}_\mathrm{m,nom}$.

The ADS is \faildegraded\ when the fault combination~$f_\mathrm{NADF}$ leads to a certain set of missions being infeasible, yielding $\mathcal{M}_\mathrm{a} \subsetneq\mathcal{M}_\mathrm{nom}$ with $\mathcal{M}_\mathrm{a}\neq\emptyset$, 
or reduces the achievable quality of the available missions $\vec{q}_\mathrm{m,a}(f_\mathrm{NADF})<\vec{q}_\mathrm{m,nom}$.
Then, the normal operation automated driving functionality is still functional, $o_\mathrm{ADS}(f_\mathrm{NADF})\,=\,o_\mathrm{NADF}(f_\mathrm{NADF})\,=\,1$, 
yet with decreased available performance, $p_{\mathrm{ADS,a}}(f_\mathrm{NADF})\,=\,p_{\mathrm{NADF,a}}(f_\mathrm{NADF})\,<\,p_\mathrm{ADS,nom}$. %
For example, faults in a redundantly designed environment perception system can lead to this system property. 
Since~${o_\mathrm{NADF}(f_\mathrm{NADF})=1}$, it is expected that the automated driving system and, thus, the automated vehicle maintain a \safestate\footnote{For this example, we presume a proof that the minimal risk condition is a \safestate. Still, providing such a proof for a given operational design domain is beyond the focus of this papers and requires further research.}.%

While \failop\ and \faildegraded\ behavior can be achieved for the example ADS within the normal operation automated driving functionality, \failsafe\ behavior of the ADS can arise in two ways. 
For both ways, the fault combination~$f_\mathrm{NADF}$ leads to all missions being infeasible, yielding $\mathcal{M}_{\mathrm{a}}~=~\emptyset$.
For the first option, the \safestate\ is maintained by the normal operation automated driving functionality through executing a minimal risk maneuver to transition the vehicle to a minimal risk condition. 
This results in $o_\mathrm{ADS}(f_\mathrm{NADF})~=~o_\mathrm{NADF}(f_\mathrm{NADF})=0$. 

The second way to implement \failsafe\ behavior is the Safe Halt functionality. 
The Safe Halt functionality can engage if the normal operation automated driving functionality is not able to maintain a \safestate\ in the presence of the fault combination $f_\mathrm{NADF}$ resulting in  $o_\mathrm{NADF}(f_\mathrm{NADF})=-1$.
By providing a minimal risk maneuver that terminates in the preselected minimal risk condition, a \safestate\ can be maintained via the Safe Halt functionality.
Therefore, $o_\mathrm{ADS}(f_\mathrm{NADF})=0$ results even for $o_\mathrm{NADF}(f_\mathrm{NADF})=-1$. 
The \failop\ trajectory selection, which is required for both options, selects the reference trajectory based on the health status of the ADS subsystems.

Overall, this example demonstrates that the taxonomy can also be applied at higher system levels and in more complex contexts. 
It supports system designers by providing the minimal required performance of a fallback system to cope with faults in the primary system to maintain a \safestate\ of the superimposed system.

\subsection{Requirements Verification}
\label{subsec:verification}

We developed our definitions for fault tolerance regimes based on the requirements given in \autoref{subsec:requirements}. 
These requirements are verified in this subsection.

Requirement~\ref{req:distinguishable} demands clear distinguishability of the fault tolerance regimes.
To address this requirement, we introduce definitions for functionality as well as available and nominal performance. 
We include these terms in the definition of fault tolerance regimes together with the scheme provided in \autoref{fig:tree}. 
Furthermore, we argue that a system can only be classified with regard to a fault tolerance regime for a specific fault combination. 
For the classification, it is necessary to assess whether a system is able to provide its specified functionality while also considering whether the system remains in a \safestate.
The nominal performance as well as the available performance due to faults under specified environmental conditions need to be identified during design time to achieve the desired fault tolerance behavior. 
This is also shown in the two examples in Subsections~\ref{subsec:steerbywire} and~\ref{subsec:safehalt}.
Within these examples, the nominal and available performance are described so that the implemented behavior of the system after the occurrence of example faults can be assigned to a fault tolerance regime.

In \autoref{sec:stateoftheart}, we show that the state-of-the-art definitions differ or are even inconsistent from each other. 
Nevertheless, as required by requirement~\ref{req:compatibility}, the proposed definitions follow the basic understanding of fault tolerance regimes in the literature. 
The proposed definitions pick up the three most commonly used terms denoting fault tolerance regimes. 
Moreover, our definitions differ from, yet are still compatible to the recently published technical report ISO/TR\,4804~\cite{iso_2020}. 

Using automotive standards as basis for the development of our definitions paves the way to applicability in the automotive domain as required by requirement~\ref{req:automotivedomain}.  
As outlined in \autoref{subsec:relatedterms}, we employ existing definitions from the automotive domain to guide the novel definitions, but also refine the \safestate\ to resolve identified contradictions.
The examples from the automotive domain, presented in Subsections~\ref{subsec:steerbywire} and~\ref{subsec:safehalt}, provide further evidence for the verification of requirement~\ref{req:automotivedomain}. 

Requirement~\ref{req:systemhierarchy} demands applicability at different system levels.
For the verification of this requirement, the two presented examples show how lower level entities can be rated by considering the implications of their behavior at higher levels. 
Additionally, the examples demonstrate that the taxonomy can be applied at system levels other than the vehicle level, which is commonly used in the automotive domain.
Thus, the taxonomy fulfills requirement~\ref{req:systemhierarchy}.

Finally, to fulfill requirement~\ref{req:multifault} to address multiple concurrent faults, we provide definitions for fault tolerance regimes covering systems' reactions to \emph{fault combinations}.
Fault combinations can consist of a single or multiple faults. 

In conclusion, we are able to verify the requirements given in \autoref{subsec:requirements} by complying with the constraints to commonly accepted definitions, adding the consideration of functionality and performance to our definition of fault tolerance regimes, and consequently applying our taxonomy to two different examples.

\section{CONCLUSION}
In this paper, we present a taxonomy to clearly distinguish fault tolerance regimes in order to overcome a diverging use found in the literature related to the automotive domain. 
The application of the taxonomy to different automotive systems~-- two examples are outlined in this paper~-- indicates its general applicability. 
To this end, the taxonomy presumes a clear definition of the system's functionality and according performance metrics.
The taxonomy and the derived definitions are compatible to the recently published technical report ISO/TR\,4804~\cite{iso_2020}, which defines \failop, \faildegraded, and \failsafe. 
Still, our taxonomy allows for an application at arbitrary system levels while ISO/TR\,4804 is limited to automated driving systems at the vehicle level.
Moreover, the taxonomy integrates into recent automotive E/E safety standards, where the focus of this paper is on \emph{functional safety} according to ISO\,26262~\cite{iso_2018}. 
The defined terms mostly refer to definitions stemming from ISO\,26262.
However, we present a concise definition of the \safestate\ because of inconsistent definitions in ISO\,26262, ISO/DIS\,21448~\cite{iso_2021}, and ISO/TR\,4804.

In general, \emph{fault tolerance regimes} can be used either ex ante to define system requirements or ex post when evaluating a system's fault tolerance.
Furthermore, we recognize the potential of our definitions to also work for domains other than the automotive domain, although we rely primarily on terms from automotive standards as a basis for our taxonomy~(\cf\ \autoref{subsec:relatedterms}). 
The automotive standards are in general compatible with more generic technical standards (\eg, IEC\,61508~\cite{iec_2010}). 
Additionally, the presented example for a steer-by-wire system may also be applied to any other redundant actuation system of another technical domain, wherein torque and actuated way are relevant factors to fulfill the specified functionality.
	
Finally, we would like to point out that the taxonomy presented by us is pivotally based on a harmonization of related works and standards and, therefore, has primarily a theoretical foundation.
The validation of the applicability of our definitions in industrial series development of a safety-critical system is still pending.
In particular, it could be questioned whether the discretization of faults and fault combinations, which is implicitly introduced by the definitions of fault tolerance regimes, is actually practicable in real complex systems.%

Our future work targets an extension of the taxonomy towards \emph{safety of the intended functionality} according to ISO/DIS\,21448~\cite{iso_2021}.  
Finally, an in-depth discussion towards linking the understanding of safety in the engineering and the legal domain is necessary, in particular for automated driving.

\section*{ACKNOWLEDGMENT}
The authors would like to thank Moritz Lippert and Tom Michael Gasser for discussing the contents of this paper as well as Sonja Luther and Ibrahim Khan for proofreading.

\renewcommand*{\bibfont}{\footnotesize} 
\printbibliography %

\begin{IEEEbiography}[]{Torben Stolte}
	studied Automation Technologies at Universität Lüneburg (Diplom (FH) 2008) and Electrical Engineering at Technische Universität Braunschweig (M.\,Sc. 2011). 
Since 2011 he is a research assistant at the Institute of Control Engineering of Technische Universität Braunschweig. 
Parallely, he worked in a collaboration with Porsche Engineering as functional safety engineer from 2011 to 2014. 
His research interest is safety of automated vehicles. He investigates the potential of fault-tolerant vehicle motion control towards safety of steering, brake, and drive actuators. 

\end{IEEEbiography}

\begin{IEEEbiography}[]{Stefan Ackermann}
	finished his Bachelor of Science and Master of Science Degrees in Mechatronics in the Field of Automotive Mechatronics at Technical University of Darmstadt. Since 2017 he is a research assistant at the Institute of Automotive Engineering at Technical University of Darmstadt. His research focuses on safety applications for automated vehicles. While pursuing his PhD he develops fallback systems for the dynamic driving task of automated vehicles.

\end{IEEEbiography}

\enlargethispage{-.25cm}\vfill

\begin{IEEEbiography}[]{Robert Graubohm}
	is a research assistant with the Institute of Control Engineering at the Technische Universität Braunschweig. Before that, he finished a Master of Science Degree in Industrial Engineering in the Field Mechanical Engineering at TU Braunschweig and a Master of Business Administration at the University of Rhode Island. His main research interests are development processes of automated driving functions and the safety conception in an early design stage.

\end{IEEEbiography}

\begin{IEEEbiography}[]{Inga Jatzkowski}
	works as a research assistant at the Institute of Control Engineering at TU Braunschweig since 2016 and is currently pursuing her PhD. She holds a Master of Science Degree in Navigation and Field Robotics from Leibniz University Hannover. Her main research topics are self-awareness and the development of self-perception for automated vehicles.

\end{IEEEbiography}

\begin{IEEEbiography}[]{Björn Klamann}
	finished his Master of Science Degree in Mechanical and Process Engineering at Technical University of Darmstadt. Since 2018 he is a research assistant at the Institute of Automotive Engineering at Technical University of Darmstadt. In his main research topic, the safety of automated vehicles, he investigates the approach of a modular safety approval.

\end{IEEEbiography}%

\begin{IEEEbiography}[]{Hermann Winner}
		began working at Robert Bosch GmbH in 1987, after receiving his PhD in physics, focusing on the predevelopment of “by-wire” technology and Adaptive Cruise Control (ACC). 
	Beginning in 1995, he led the series development of ACC up to the start of production. 
	Since 2002 he has been pursuing the research of systems engineering topics for driver assistance systems and automated driving as Professor of Automotive Engineering at the Technische Universit\"at Darmstadt. 
	He discovered the ``approval trap'' of autonomous driving, the still unsolved challenge to validate safety of autonomous driving before market introduction.

\end{IEEEbiography}%

\begin{IEEEbiography}[]{Markus Maurer}
	  received the Diploma degree in electrical engineering from the Technische Universität München, in 1993, and the Ph.D. degree in automated driving from the Group of Prof. E. D. Dickmanns, Universität der Bundeswehr München, in 2000.
 From 1999 to 2007, he was a Project Manager and the Head of the Development Department of Driver Assistance Systems, Audi.
 Since 2007, he has been a Full Professor of automotive electronics systems with the Institute of Control Engineering, Technische Universität Braunschweig.
 His research interests include both functional and systemic aspects of automated road vehicles.

\end{IEEEbiography}

\end{document}